\documentclass[twoside, twocolumn, english, superscriptaddress,showpacs, longbibliography]{revtex4-2}
\usepackage[T1]{fontenc}
\usepackage[utf8]{inputenc}
\usepackage[english]{babel}
\usepackage{bm}
\usepackage{amsmath}
\usepackage{amssymb}
\usepackage{graphicx}
\usepackage[unicode=true, pdfusetitle, bookmarks=true,bookmarksnumbered=true,bookmarksopen=true,bookmarksopenlevel=1,
breaklinks=false,pdfborder={0 0 0},pdfborderstyle={},backref=false,color links, citecolor=blue,linkcolor=red,urlcolor=blue] {hyperref}
\usepackage[nameinlink]{cleveref}
\Crefname{figure}{Fig.}{}

\makeatletter
\usepackage{braket}
\usepackage{bm}
\usepackage{tikz}
\usepackage{pgfplots}
\usepackage{pgf}
\usepackage{subfigure}
\usepackage{nicematrix}
\usepackage{diagbox}
\usepackage{orcidlink}

\renewcommand{\selectlanguage}[1]{}

\makeatother

\begin{document}
\title{Universal Modelling of Oscillations in Fractional Quantum Hall Fluids}
\author{Guangyue Ji}
\affiliation{Division of Physics and Applied Physics, Nanyang Technological University, Singapore 637371}

\author{Koyena Bose}
\affiliation{Institute of Mathematical Sciences, CIT Campus, Chennai 600113, India}
\affiliation{Homi Bhabha National Institute, Training School Complex, Anushaktinagar, Mumbai 400094, India}

\author{Ajit C. Balram\orcidlink{0000-0002-8087-6015}}
\affiliation{Institute of Mathematical Sciences, CIT Campus, Chennai 600113, India}
\affiliation{Homi Bhabha National Institute, Training School Complex, Anushaktinagar, Mumbai 400094, India}

\author{Bo Yang}
\affiliation{Division of Physics and Applied Physics, Nanyang Technological University,
Singapore 637371}

\begin{abstract}
Density oscillations in quantum fluids can reveal their fundamental characteristic features. In this work, we study the density oscillation of incompressible fractional quantum Hall (FQH) fluids created by flux insertion. For the model Laughlin state, we find that the complex oscillations seen in various density profiles in real space can be universally captured by a simple damped oscillator model in the occupation-number space. It requires only two independent fitting parameters or characteristic length scales: the decay length and the oscillation wave number. Realistic Coulomb quasiholes can be viewed as Laughlin quasiholes dressed by magnetorotons, which can be modeled by a generalized damped oscillator model. Our work reveals the fundamental connections between the oscillations seen in various aspects of FQH fluids such as in the density of quasiholes, edge, and the pair correlation function. The presented model is useful for the study of quasihole sizes for their control and braiding in experiments and large-scale numerical computation of variational energies.
\end{abstract}

\maketitle

\section{Introduction}
\label{sec:introduction}
Characteristic density oscillations in the presence of perturbations are a fundamental aspect of quantum fluids. For the Fermi liquid, the long-range Friedel oscillation in the presence of impurity is a direct consequence of the existence of the Fermi surface and quasiparticle excitations~\citep{giuliani2008quantum}. Similarly, for the non-Fermi liquids and strongly interacting topological fluids which go beyond the Fermi liquid paradigm (e.g. the Luttinger liquid~\citep{haldane1981luttinger}, composite fermion liquid~\citep{halperin_theory_1993}, quantum Hall liquid~\citep{klitzing_new_1980,tsui_two-dimensional_1982} and strange metal~\citep{anderson_strange_2006}), characteristic density oscillations manifest as spin-charge separation~\citep{senaratne_spin-charge_2022}, charge-vortex duality~\citep{halperin2020half}, anomalous decay laws and exponents~\citep{Kamilla97, chang_chiral_2003, balram_luttinger_2015, balram_luttinger_2017, mitrano2018anomalous}, and so on. Understanding and accurately modeling such oscillations, which are of both theoretical and experimental significance, remains an outstanding open problem.

\begin{figure}[t]
\includegraphics[width=1\columnwidth, height=0.5\columnwidth]{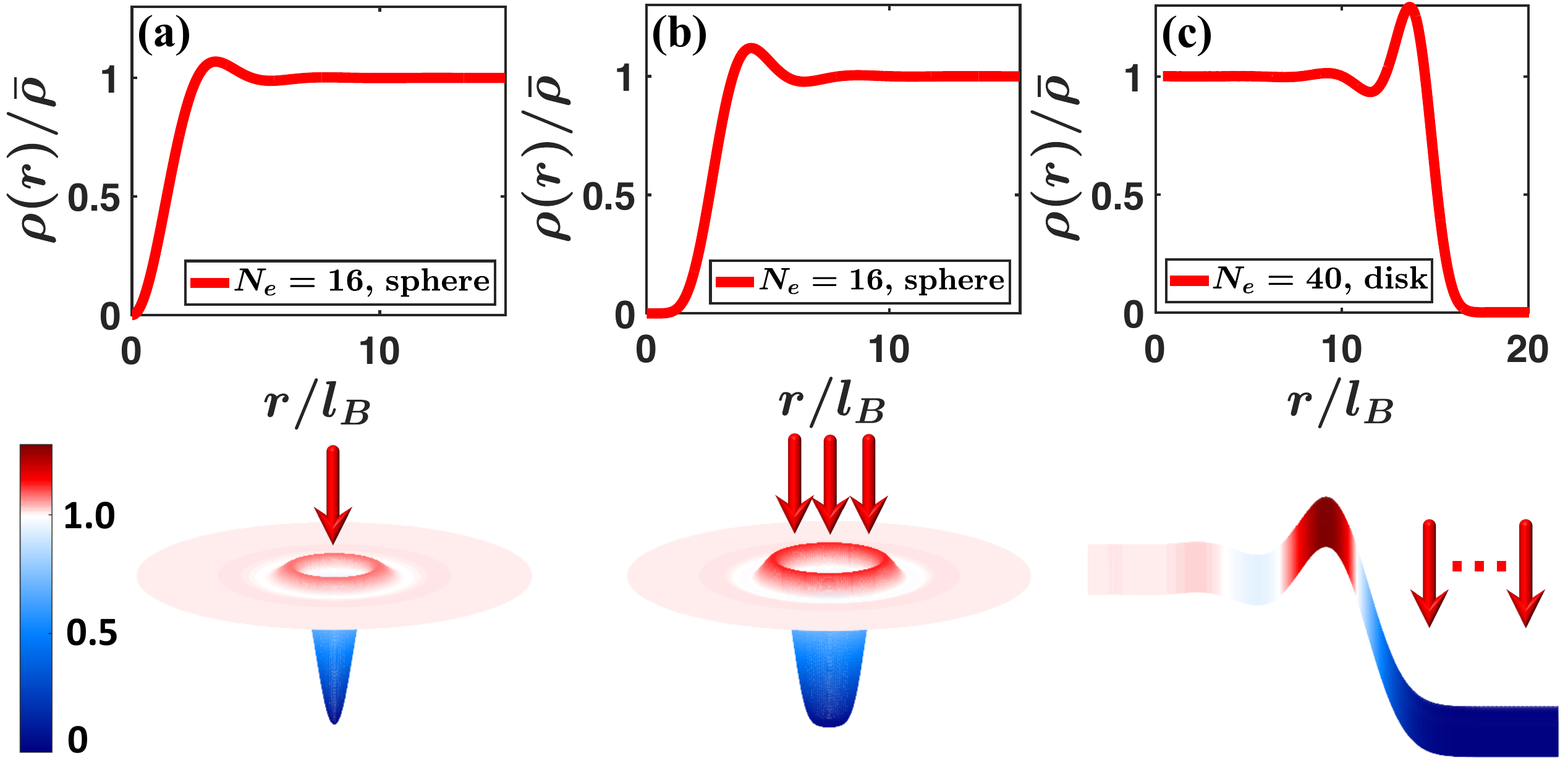}
\caption{\label{fig: schematic_plot} Schematic plot of excitations created by flux insertion at $\nu{=}1/3$. From left to right, the Laughlin quasihole obtained by insertion of a single flux (a), $3$-Laughlin quasiholes obtained by insertion of three fluxes (b), whose density is proportional to the ground state pair-correlation function, and the Laughlin edge obtained from the insertion of an infinite number of fluxes (c). The top panels show the density distribution $\rho(r)/\bar{\rho}$ in the radial direction, where $\bar{\rho}$ is the uniform background density. Panels (a) and (b) show results for $N_{e}{=}16$ on the sphere while panel (c) is for $N_{e}{=}40$ on the disk. 
}
\end{figure}

An important class of strongly interacting topological fluids is the incompressible fractional quantum Hall (FQH) fluids which display characteristic oscillatory features that encode both geometric and universal topological information. When a quantum of flux is inserted into the uniform FQH ground state, quasiholes carrying a fractional charge~\citep{laughlin_anomalous_1983} and obeying fractional statistics~\citep{arovas_fractional_1984} are created. They also carry a dipole moment to balance the Hall viscosity in the presence of the electric field gradient, which is proportional to the FQH topological shift~\citep{trung_spin-statistics_2023}. The dipole moment, a characteristic feature of incompressibility, is established from the density oscillation at the edge~\citep{park_guiding-center_2014}. Moreover, when an appropriate number of fluxes equivalent to removing an electron is inserted at the same position, the density of the bound state of the stacked quasiholes is proportional to the pair-correlation function of the ground state~\citep{datta_edge_1996}. In the limit of an infinite number of fluxes inserted, a macroscopic FQH edge is created, near which the density oscillation has been intensively studied~\citep{wen_chiral_1990, datta_edge_1996, levesque2000charge, wiegmann_nonlinear_2012, park_guiding-center_2014, can_singular_2014, fern_quantum_2017, ito_density_2021, cardoso_boundary_2021, yang_monte_2023}. The FQH quasiholes, ground state pair-correlation function, and the edge can be understood as special cases of \emph{flux insertion} (see Fig.~\ref{fig: schematic_plot}) and are thus closely related. The precise underlying connections and both the qualitative and quantitative aspects of such oscillations, however, are not well understood.

In this paper, we show that the \emph{real space density oscillation} from the flux insertion in the model Laughlin state can be accurately modeled by a simple damped oscillation with degrees of freedom within a single Landau level (LL). The oscillation is determined by two characteristic length scales: the decay length and the oscillation wave number, and these serve as the only fitting parameters of the model. In contrast to previous works that directly model the real space density~\citep{girvin_magneto-roton_1986, levesque2000charge, wiegmann_nonlinear_2012, johri_quasiholes_2014,can_singular_2014, balram_luttinger_2015, balram_luttinger_2017, cardoso_boundary_2021, fulsebakke_parametrization_2023}, we emphasize that the model should only focus on the guiding center degrees of freedom (within a single LL), as those are the relevant coordinates for any FQH phase. For the model Laughlin state, we study its quasihole, edge, and pair-correlation function using the aforementioned model-based approach treating them all on an equal footing. A phenomenological model for the damped oscillation of Laughlin quasiholes is proposed at general fillings. For the more realistic Coulomb interaction, where quasiholes are dressed by neutral excitations, a generalized damping model with four characteristic lengths is shown to work very well. This general model can be useful for both numerical computations and experimental manipulation of quasiholes. Moreover, we find this generalized model also gives an accurate description of quasiholes in some of the composite fermion and non-Abelian topological phases.

The remainder of the paper is organized as follows. In Sec.~\ref{sec:Universal_damped_oscillation}, we study the density oscillation of a single Laughlin quasihole. In Sec.~\ref{sec:Stacked_Laughlin_quasiholes}, we generalize the study to various stacked Laughlin quasiholes. Moving forward, we examine the density oscillation of quasihole for realistic interactions in Sec.~\ref{sec:Quasiholes_realistic_interactions}, and for Jain and non-Abelian phases in Sec.~\ref{sec:non_Abelian_QH}. Thereafter, we also investigate the oscillatory behaviors for quasielectrons in Sec.~\ref{sec:Quasielectron}. Finally, we summarize our results in Sec.~\ref{sec:Summary}.

\section{Universal damped oscillation}
\label{sec:Universal_damped_oscillation}
\begin{figure}[t]
\includegraphics[width=0.49\columnwidth]{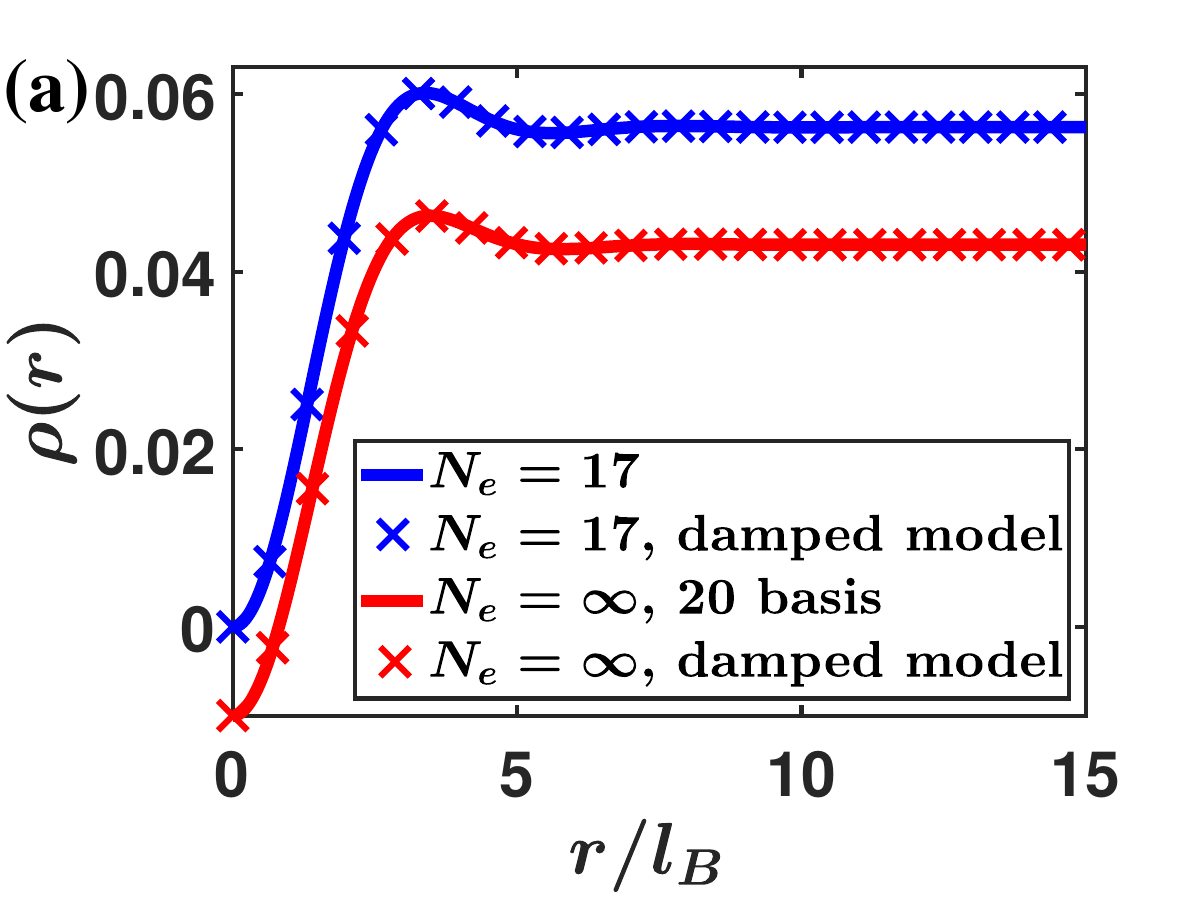}\hfill{}
\includegraphics[width=0.49\columnwidth]{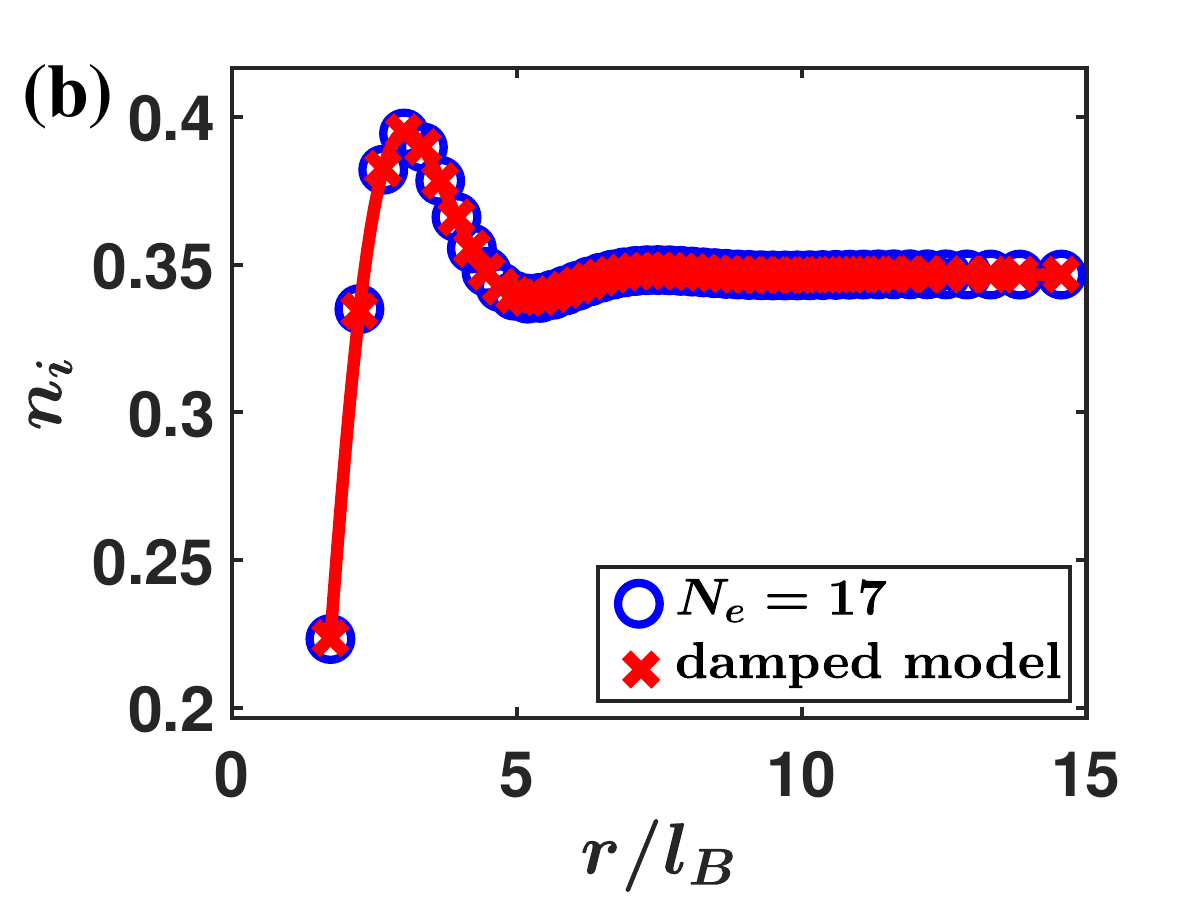}
\caption{\label{fig: single_quasihole} (a) The density distribution of a Laughlin quasihole located at the north pole of a sphere as a function of the arc distance $r$. The blue line and crosses denote the exact distribution and the fitting with our damped model [Eq.~\eqref{eq:obital_fitting}], respectively for $N_{e}{=}17$. The red line and crosses denote the thermodynamic limit obtained from the polynomial expansion method~\citep{fulsebakke_parametrization_2023} and our fitting [see Appendix.~\ref{sec:1over3_QH} for details], respectively, and the data points are shifted down by $0.01$ for clarity. (b) The occupation-number density. Blue circles and red crosses denote the exact density and that obtained from fitting with our model, respectively. The fitted curve is $n(r){=} 0.68\sin[1.43(r{-}2.31)]\exp({-}r/1.21){+}\bar{n}$ with $\bar{n}{=}(N_{e}{+}1/3)/N_{o}$ with $N_{o}{=}50$.
}
\end{figure}

The model wave function of the Laughlin quasihole located at the origin of the disk geometry for $N_{e}$ electrons is $\Psi_{\rm L}^{1{-}\text{qh}}{=}\left(\prod_{i{=}1}^{N_{e}}z_{i}\right) \Psi_{\rm L},$ where $\Psi_{\rm L}{=}\prod_{1{\leq}i{<}j{\leq}N_{e}}(z_{i}{-}z_{j})^{m}e^{-\frac{1}{4}\sum_{i{=}1}^{N_{e}}\vert z_{i}\vert^{2}}$ is the Laughlin wave function at $\nu{=}1/m$~\citep{laughlin_anomalous_1983}, $z_{i}{=}x_{i}{-}iy_{i}$ is the position of the $i^{\text{th}}$ electron, and the magnetic length at magnetic field $B$ is taken as the unit of length i.e., we set $l_{B}{=}\sqrt{\hbar c/(eB)}{=}1$. We can also stereographically map this state to the spherical geometry~\citep{haldane_fractional_1983}. The density distribution of a single Laughlin quasihole at $\nu{=}1/3$ in the spherical geometry is shown in Fig.~\ref{fig: single_quasihole}(a). Directly finding a simple empirical model for the real-space density distribution of quasiholes is challenging as the real-space structure is a mixture of the trivial LL (cyclotron) and the non-trivial FQH (guiding-center) contributions. A real-space density distribution $\rho(r)$ can be decomposed as $\rho (r) {=}\sum_{i{=}0}^{N_{\phi}} n_{i} \rho_{i}(r)$, where $n_{i}$ is the average occupation number of the $i^{\text{th}}$ orbital, $\rho_i$ is the density computed from single-particle wave functions, and $N_{\phi}$ is the number of fluxes threading the sample. Only $n_{i}$ is related to the correlated FQH physics. 

The Laughlin quasihole density exhibits damped oscillations in the occupation-number space [see Fig.~\ref{fig: single_quasihole}(b)]. We take the real space position of each orbital as the arc distance from the north pole to the center of the $i^{\text{th}}$ equal-area slice of the sphere, i.e., $x_{i}{=}R\theta_{i}{=}\sqrt{N_{\phi}/2}\arccos\left[1{-}\left(1{+}2i\right)/N_{o}\right],$ where $R{=}\sqrt{N_{\phi}/2}$ is the radius of the sphere, $N_{o}{=}N_{\phi}{+}1$ is the number of orbitals on the sphere, and we index the orbitals from the north to the south pole by $k{=}0,1,\ldots, N_{\phi}$. We find the occupation-number oscillation of quasihole density $\delta n_{i}{=}n_{i}{-}\bar{n}$ [with $\bar{n}{=}(N_{e}{+}1/m)/N_{o}$  being the average background occupation] can be accurately captured by the following model:
\begin{equation}
\delta n_{i} =A_{1}\sin[k_{1}(x_{i}-x_{1})]\exp\left(-x_{i}/\lambda_{1}\right), 
\label{eq:obital_fitting}
\end{equation}
where $k_{1}$ and $\lambda_{1}$ are the oscillation wave number and decay length respectively, $A_{1}$ is the amplitude and $x_{1}$ is the zero point. The fitted result (red crosses) agrees almost perfectly with the exact occupations (blue circles) as shown in Fig.~\ref{fig: single_quasihole}(b). Correspondingly, the exact and fitted real-space density distributions are also nearly indistinguishable from each other as shown in Fig.~\ref{fig: single_quasihole}(a).

It is important to note that the accurate fitting is achieved with only two independent fitting parameters $k_{1}$ and $\lambda_{1}$. The other two parameters in Eq.~\eqref{eq:obital_fitting}, $A_1$ and $x_1$, are fixed by the total charge $\sum_{i{=}0}^{N_{\phi}} \nu_{i}{=}N_{e}$ and the total angular momentum $\sum_{i{=}0}^{N_{\phi}}\nu_{i}(i{-}N_{\phi}/2){=}N_{e}/2$. The two conditions are equivalent to the constraints that a single quasihole has a charge $e_{1{-}\text{qh}}{=}{-}1/m$ and a dipole moment $d_{1{-}\text{qh}}{=}(1{-}1/m)/2$ in the thermodynamic limit. Both of these relations are topological and the focus of $K$-matrix Luttinger liquid theory, thus robust against perturbations~\citep{wen_chiral_1990}. From finite-size-scaling, we find the following values of the parameters in the thermodynamic limit [see Appendix.~\ref{sec:1over3_QH} for details]: 
\begin{equation}
k_{1}^{(1)}=1.38, \lambda_{1}^{(1)} = 1.17, \label{eq:orbital_parameters}
\end{equation}
where the superscript denotes the number of quasiholes. In contrast, if one works with the real-space density which includes the extra Landau orbit contribution, one has to introduce several tens of fitting parameters using the polynomial expansion method to capture the whole profile as done in Refs.~\citep{girvin_magneto-roton_1986, fulsebakke_parametrization_2023}. Furthermore, the important property that the Laughlin quasihole has a unique decay length and oscillation period is not easy to discern from real-space studies ~\citep{johri_quasiholes_2014}. 

\section{Stacked Laughlin quasiholes}
\label{sec:Stacked_Laughlin_quasiholes}
\begin{figure}[t]
\includegraphics[width=0.48\columnwidth]{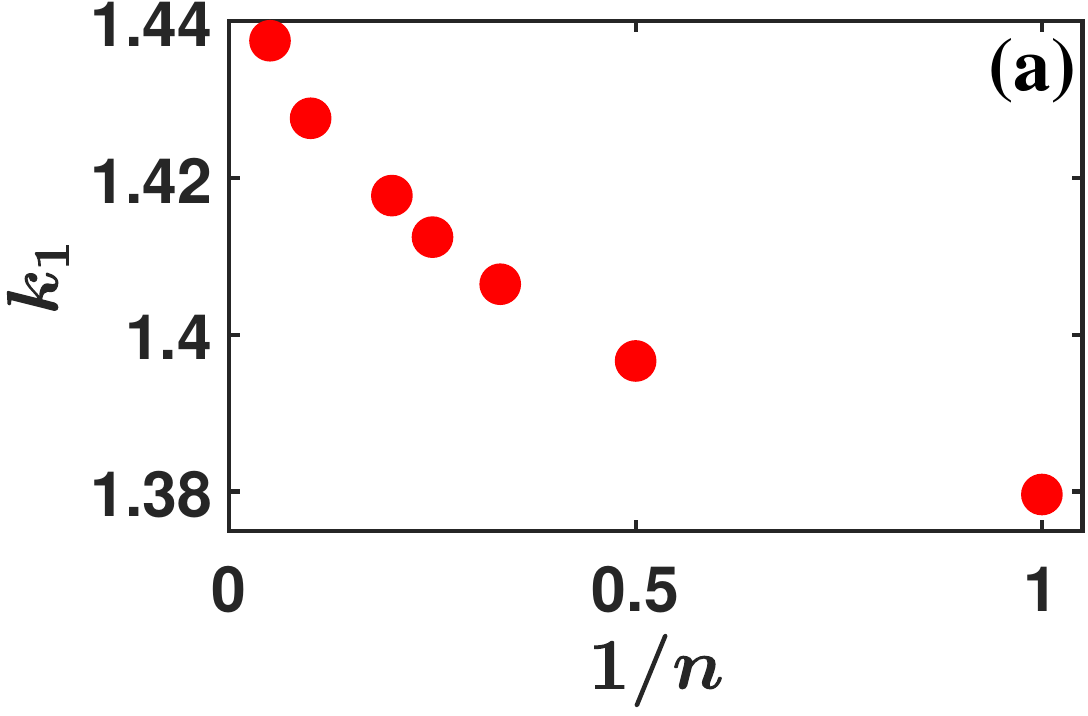}\hfill{}\includegraphics[width=0.48\columnwidth]{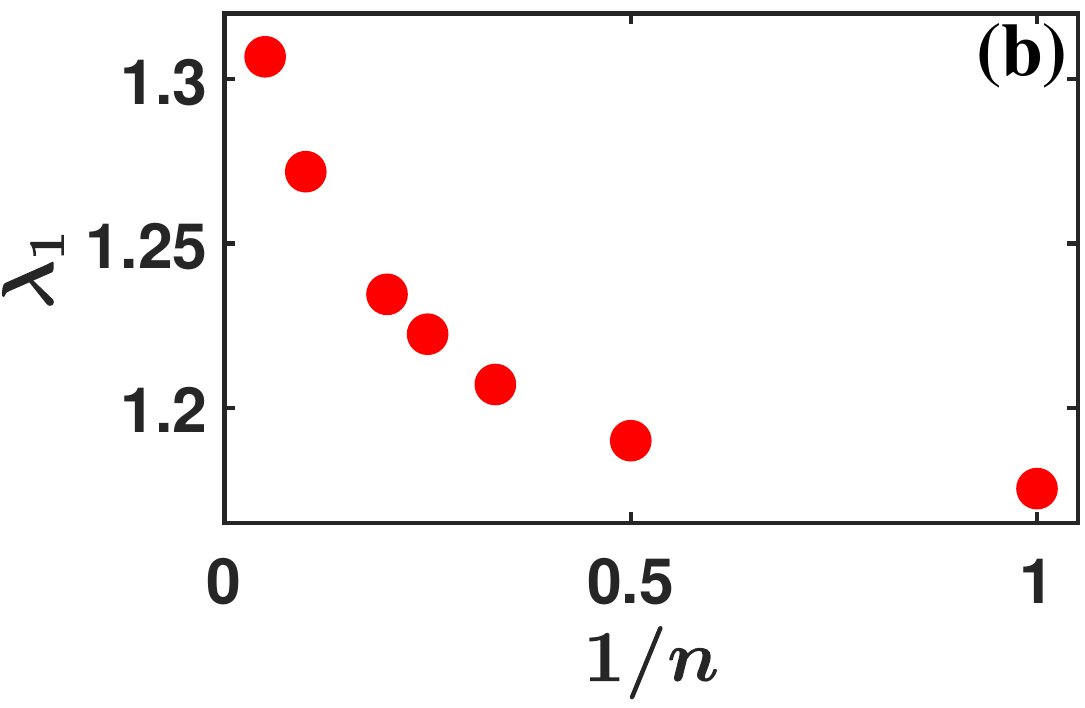}

\hspace{0.2cm}\includegraphics[width=0.45\columnwidth]{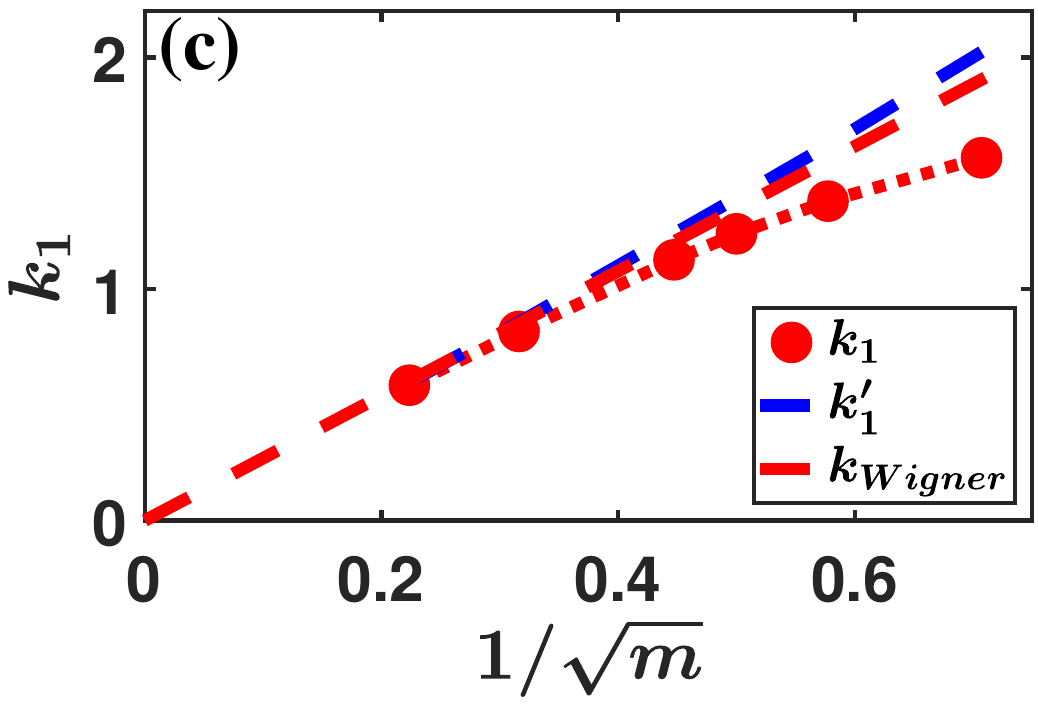}\hspace{0.3cm}
\includegraphics[width=0.48\columnwidth]{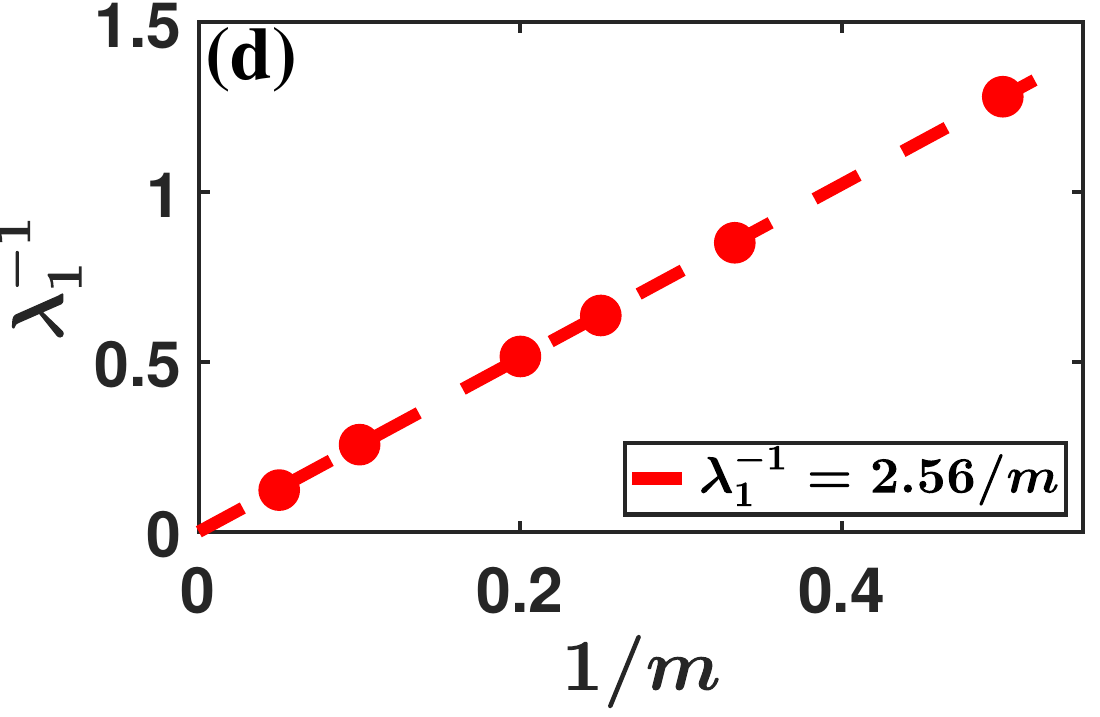}
\caption{\label{fig: characteric_length_nstacked} Top panels (a) and (b) show the characteristic lengths $k_{1}$ and $\lambda_{1}$ of Laughlin $n$-quasihole state at $1/3$. Bottom panels (c) and (d) show results for a single Laughlin quasihole at different fillings $\nu{=}1/m$. The red dashed line in (c) is the wave number of the Wigner crystal at the corresponding filling, and the blue dashed line denotes the intrinsic wave number $k^{\prime}_{1}{=}\sqrt{k_{1}^{2}{+}\lambda_{1}^{{-}2}}$ (see text).
}
\end{figure}

The damped oscillation model can also be applied to $n$-stacked quasiholes created by inserting $n$ fluxes at the same location. The real-space density of such quasiholes gets increasingly complex with increasing $n$. Even for IQH fluid, the $n$-stacked holes have $\delta \rho(r) {=} {-}1/(2\pi)\exp\left({-}r^2/2  \right) (\sum_{i{=}0}^{n{-}1} r^{2i}/2^{i} i! )$, which is no longer a simple Gaussian distribution for $n{>}1$. Focusing on the FQH system with multiple quasiholes, we again look at its occupation-number density which reveals interesting physics and is no more complex even for $n{>}1$. The values of the characteristic lengths $k_{1}^{(n)}$ and $\lambda_{1}^{(n)}$ for different values of $n$ are shown in Figs.~\ref{fig: characteric_length_nstacked}(a) and \ref{fig: characteric_length_nstacked}(b), where both $k_{1}^{(n)}$ and $\lambda_{1}^{(n)}$ increase as $n$ gets larger. Furthermore, the complicated real-space densities can be easily restored with the single-particle wave functions.

The $m$-stacked quasihole at $\nu{=}1/m$ deserves special attention because its density distribution is proportional to the pair-correlation function of the Laughlin ground state. The density distribution of $m$-Laughlin quasiholes is $\rho(r){=}\mathcal{N}_{1}\int\prod_{i{=}2}^{N_{e}^{\prime}}d\bm{r}_{i}\vert\Psi_{L}^{m{-}\text{qh}}(\bm{r},\bm{r}_{2},{\ldots},\bm{r}_{N_{e}^{\prime}})\vert^{2},$ where $\mathcal{N}_{1}$ is a constant; and the pair-correlation function is $g(r){=}\mathcal{N}_{2}\int\prod_{i{=}3}^{N_{e}}d\bm{r}_{i}\vert\Psi_{L}(0,\bm{r},\bm{r}_{3},{\ldots},\bm{r}_{N_{e}})\vert^{2}.$ By substituting the explicit wave function $\Psi_{L}$ and $\Psi_{L}^{m{-}\text{qh}}$ into the expression of $g(r)$ and $\rho(r)$, respectively, one can show that $\rho(r)$ is proportional to $g(r)$ when $N_{e}{=}N_{e}^{\prime}{+}1$. Therefore, the decay length of the density of $m$-stacked Laughlin quasiholes is equal to the correlation length $\lambda_{\text{cor}}$ between electrons: 
\begin{equation}
\lambda_{\text{cor}}=\lambda^{(m)}_{1}.\label{eq:correlation_decay}
\end{equation}

Our result is thus useful for the large-scale numerical computation of ground-state variational energies. One can use our modeled $g(r)$ to calculate the per-particle variational energy of the Laughlin state in terms of any general interaction $V(r)$ through the formula $V{=}(\bar{\rho}/2) \int d^{2} \, \bm{r} V(r)[g(r){-}1]$~\citep{jain2007composite}, where $\bar{\rho}$ is the average density. For the Coulomb interaction $V(r){=}1/r$, the calculated energy $V$ with our modeled $g(r)$ is $-0.4096$ for $\nu{=}1/3$ and $-0.3278$ for $\nu{=}1/5$~[see Appendices.~\ref{sec:1over3_pair} and \ref{sec:1over5_pair}], which are very close to the conjectured thermodynamic values $-0.4098$~\citep{Ciftja03, balram_fractional_2020} and $-0.3275(1)$~\citep{dora2023competition} obtained from the extrapolation of small-size exact diagonalization results and large-scale Monte Carlo calculations. Thus for other realistic interactions, the thermodynamic variational energies can now be computed very efficiently with our modeled $g(r)$~[see Appendices.~\ref{sec:1over3_pair} and \ref{sec:1over5_pair}]. Besides, our method does not suffer from the systemic error of the polynomial expansion method at small $r$~\citep{fulsebakke_parametrization_2023}.

In the limit of $n{\rightarrow}\infty$, the corresponding ``Laughlin quasihole'' also deserves special attention as it represents the edge of the FQH fluid. Its density profile can also be well-fitted with a damped model in the occupation-number space with the characteristic lengths being $k_{1}^{\text{edge}} {=}1.49$ and $\lambda_{1}^{\text{edge}}{=}1.51$~[see Appendix.~\ref{sec:1over3_edge}] which are just the limiting values of $n$-quasiholes as shown in Fig.~\ref{fig: characteric_length_nstacked}. 

Moreover, we can determine the characteristic length scales at general Laughlin fillings $\nu{=}1/m$ by fitting the density profile of a single quasihole with the damped model given in Eq.~(\ref{eq:obital_fitting}). The results are shown in Figs.~\ref{fig: characteric_length_nstacked}(c) and \ref{fig: characteric_length_nstacked}(d). With increasing $m$, the oscillations become more pronounced. We find that the decay length is proportional to $m$, i.e., $\lambda_{1}^{(1)}{=}m/2.56$. Meanwhile, the oscillation wave number $k_{1}$ gradually approaches that of the Wigner crystal~\citep{Wigner34}, i.e., $k_{\text{Wigner}}{=}2{\times}3^{{-}1/4}\sqrt{\pi m}{=}2.69\sqrt{m}$. This is consistent with a recent study, carried out in Ref.~\citep{cardoso_boundary_2021}, of the density oscillations at the edge of the Laughlin state.

Given the near-perfect fitting of such a simple model with only two fitting parameters~[see Appendices.~\ref{sec:1over3_QH}-\ref{sec:general_laughlin}], we conjecture that the damped oscillation model captures the main features of the universal oscillations in the Laughlin FQH fluid, and in principle can be derived. While we are not able to accomplish that here, a phenomenological model for Laughlin quasiholes at general fillings can be proposed based on these results. We view the system as a damped oscillator with wave number $k^{\prime}_{1}{=}\sqrt{k_{1}^{2}{+}\lambda_{1}^{-2}}$ serving as the intrinsic ``wave number'' which is a little larger than the wave number of the Wigner crystal $k_{\text{Wigner}}$ for small $m$ [see Fig.~\ref{fig: characteric_length_nstacked}(c)]. The damping results from a ``frictional force'' which is proportional to the density $\nu{=}1/m$. The differential equation that describes this system is 
\begin{equation}
    \left[\partial_{r}^{2}+2\lambda_{1}^{-1}\partial_{r}+k^{\prime2}_{1}\right] \delta n(r) + \mathcal{O}(\delta n^2) =0,
    \label{eq:damped_oscillator_picture}
\end{equation}
where $\mathcal{O}(\delta n^2)$ denote terms that can potentially arise from non-linear effects which we ignore here. The linear solution of Eq.~\eqref{eq:damped_oscillator_picture} is just our damped model given in Eq.~(\ref{eq:obital_fitting}), which is known to model the classical two-dimensional shallow-water wave near a specified moving boundary~\citep{thacker_exact_1981,sampson2008some}. Note that the model proposed in Eq.~\eqref{eq:damped_oscillator_picture} is different from that proposed in Eq.~(23) of Ref.~\citep{wiegmann_nonlinear_2012}, which does not consider the damped term and assumes a different intrinsic wave number at the level of linear response. Deriving this simple model from the microscopic wave function and capturing the non-linear effects remains an open problem~\citep{wiegmann_nonlinear_2012,can_singular_2014}. 

\section{Quasiholes for realistic interactions}
\label{sec:Quasiholes_realistic_interactions}
\begin{figure}[t]
\includegraphics[width=0.49\columnwidth]{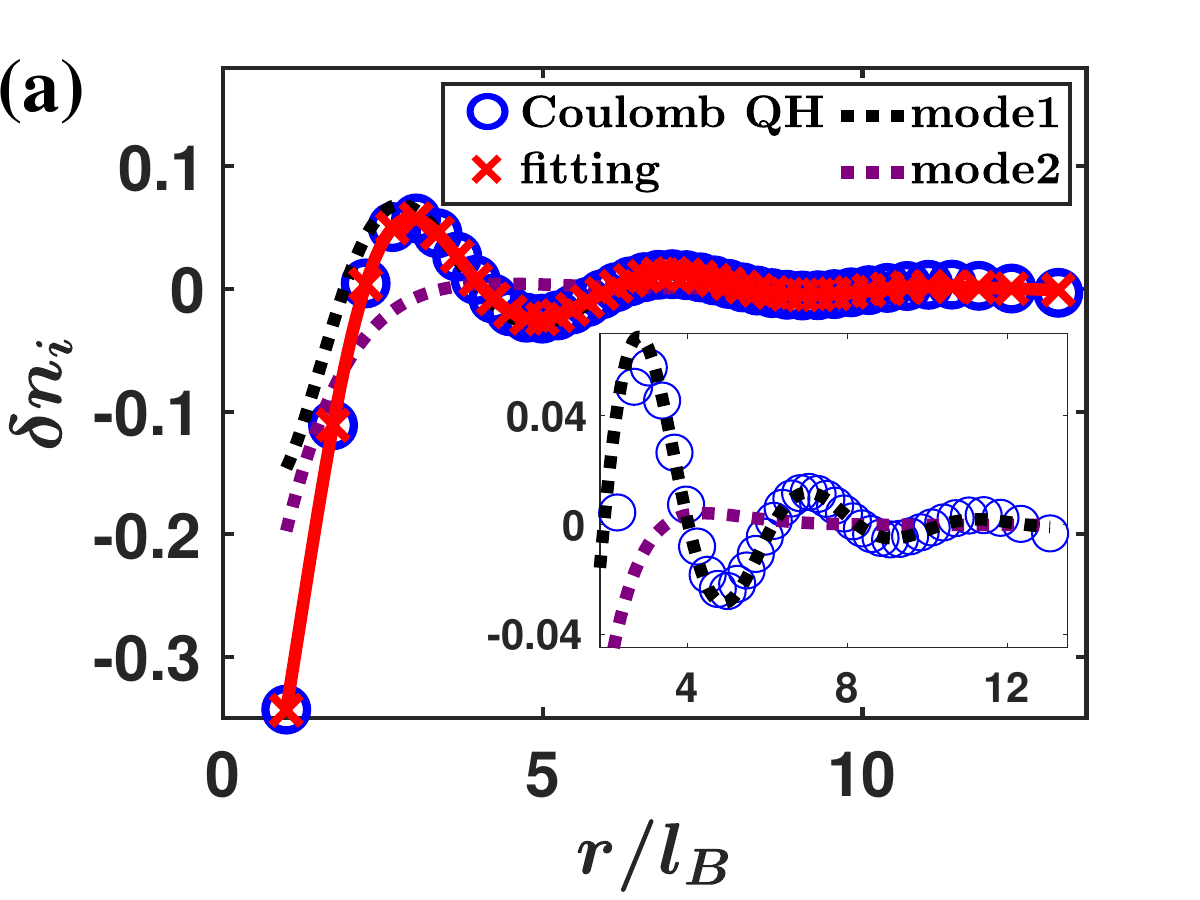}\hfill{}\includegraphics[width=0.49\columnwidth]{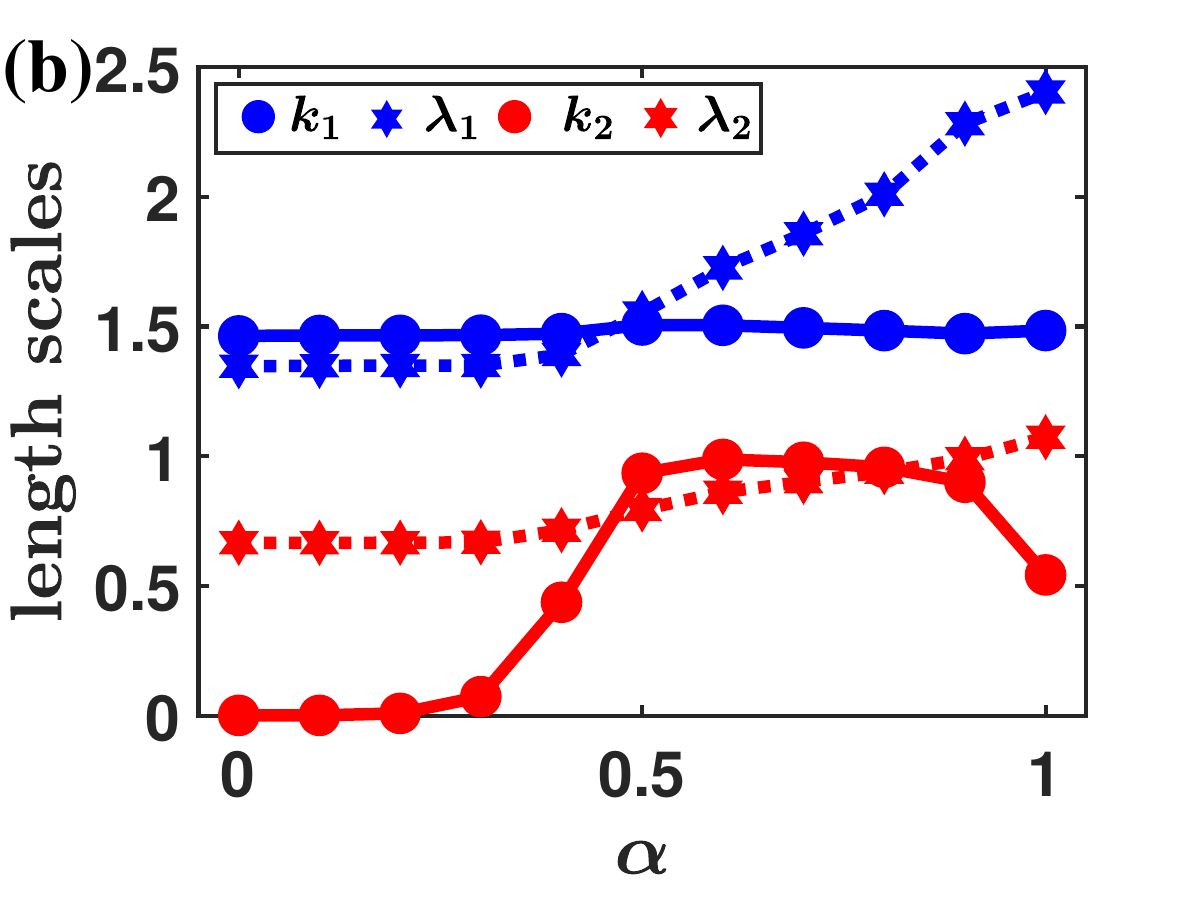}
\caption{\label{fig: result_alpha}(a) Occupations of the Coulomb quasihole at $\nu{=}1/3$ for $N_{e}{=}14$ on the sphere. The fitted curve (red line) is the sum of the long-range $0.23\sin[1.48r{-}2.82]\exp({-}0.42r)$ (black dotted line) and the short-range $0.51\sin[0.54r{-}1.86]\exp({-}0.93r)$ (purple dotted line) modes. The inset is a zoom-in on the tail of the damped oscillations. (b) The dependence of characteristic lengths of the two modes on $\alpha$ [defined in Eq.~\eqref{eq: interpolating_V1_Coulomb_H}] for $N_{e}{=}14$. The blue and red colors denote the long-range and short-range modes respectively.
}
\end{figure}

Going beyond the model Laughlin quasihole wave functions, the density profile of quasiholes from realistic interactions becomes more complicated, because such quasiholes are dressed by neutral excitations~\citep{balram_role_2013, Balram13}. Here we study the quasihole in the presence of the following tunable interaction
\begin{equation}
    H_{\alpha}=(1-\alpha) V_{1} + \alpha V_{\text{Coulomb}},
    \label{eq: interpolating_V1_Coulomb_H}
\end{equation}
where $V_{\text{Coulomb}}{=}1/r$ is the Coulomb interaction and $\alpha {\in} [0,1]$ is a tunable parameter. By varying $\alpha$, one can trace the evolution between the Laughlin quasihole ($\alpha{=}0$) and the Coulomb quasihole ($\alpha{=}1$). As $\alpha$ increases, the oscillations in the density of the quasihole become more pronounced and multiple periods/frequencies are observed. As a result, the single-damped model of Eq.~\eqref{eq:obital_fitting} can no longer capture the whole density profile of the Coulomb quasihole. Instead, we need to generalize the model to include an additional damped oscillation that has a shorter range as shown in Fig.~\ref{fig: result_alpha}(a). The explicit expression of the generalized model is 
\begin{equation}
\delta n_{i} = \sum_{n=1}^{2} A_{n}\sin[k_{n}(x_{i}-x_{n})]\exp\left(-x_{i}/\lambda_{n}\right). 
\label{eq:obital_fitting}
\end{equation}
The dependence of the characteristic lengths of the two modes on $\alpha$ is shown in Fig.~\ref{fig: result_alpha}(b). The wave number of the long-range mode $k_{1}$ is nearly unchanged and its decay length $\lambda_1$ becomes larger as $\alpha$ increases. This is not hard to understand, because the interaction becomes more long-ranged as $\alpha$ is increased. For the short-range mode, its wave number $k_2$ approaches $0$ as $\alpha {\rightarrow} 0$ and it provides nearly no contribution to non-empty orbitals of Laughlin quasihole. This is consistent with the result that the Laughlin quasihole can be well-modeled with a single-damped oscillation. 

Although the generalized model captures the density profile of the Coulomb quasihole accurately, one should be careful of issues about possible overfitting and strong finite size effects [see Appendix.~\ref{sec:Coulomb_quasihole} for a discussion of these]. Nevertheless, the high accuracy of the fitting suggests that the model captures at least two oscillation modes for the Coulomb quasihole. Our result quantitatively characterizes the deviation between the Coulomb and Laughlin quasiholes and reveals the non-universal effects induced by realistic interactions. This deviation arises from the dressing of the Laughlin quasihole by low-energy magnetoroton states~\citep{yang_model_2012, balram_role_2013, Balram13}. Since the Coulomb quasihole and the Coulomb edge are closely related by flux insertion, the low-energy magnetoroton mode should also account for the deviation between the Coulomb edge and the Laughlin edge as observed in Refs.~\citep{ito_density_2021, yang_monte_2023}. Moreover, other non-universal effects away from the chiral Luttinger theory~\citep{chang_chiral_2003,mandal2001universal,wan_universality_2005} can potentially be understood by studying the microscopic bulk quasihole~\citep{yang_statistical_2021} via bulk-boundary correspondence, which requires much less computational efforts because its fluctuation and related non-linear effects are much smaller compared to the edge effects.

\section{Quasiholes in Jain and non-Abelian phases}
\label{sec:non_Abelian_QH}
\begin{figure}[t]
\subfigure[Moore-Read quasihole]{\includegraphics[width=0.48\columnwidth]{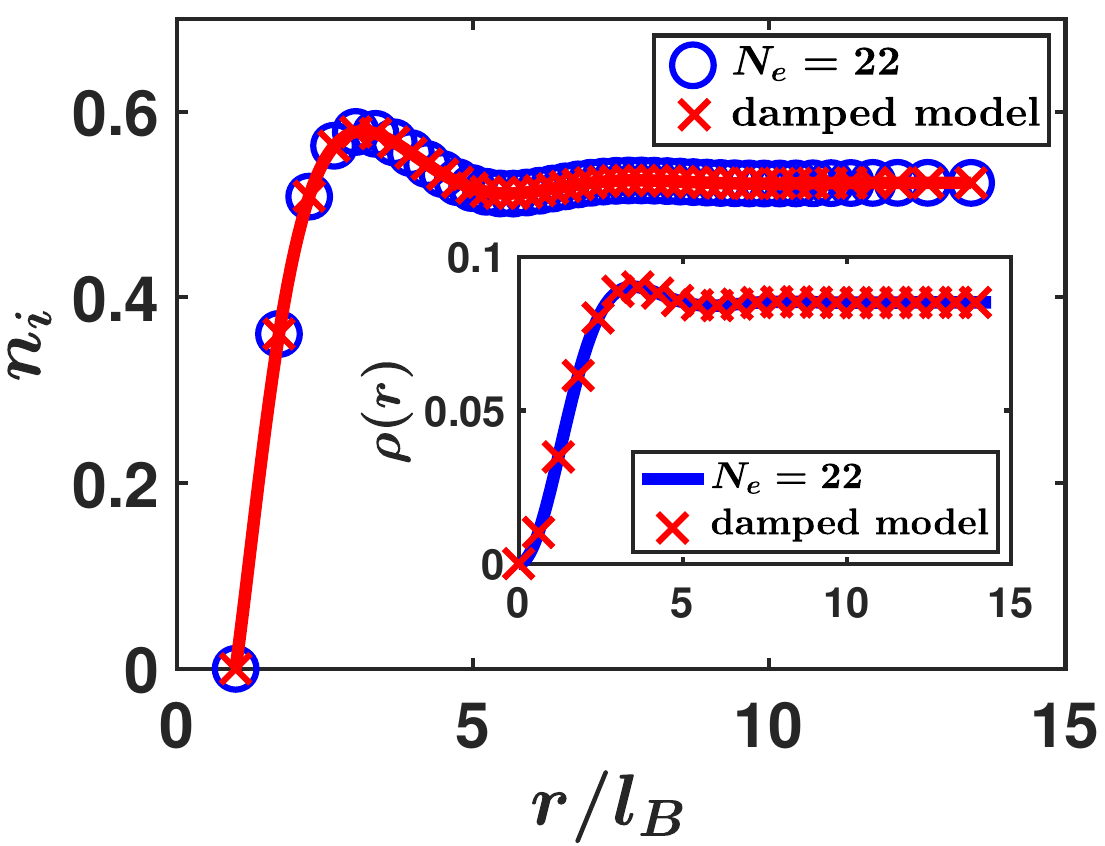}}\subfigure[Gaffnian quasihole]{\includegraphics[width=0.48\columnwidth]{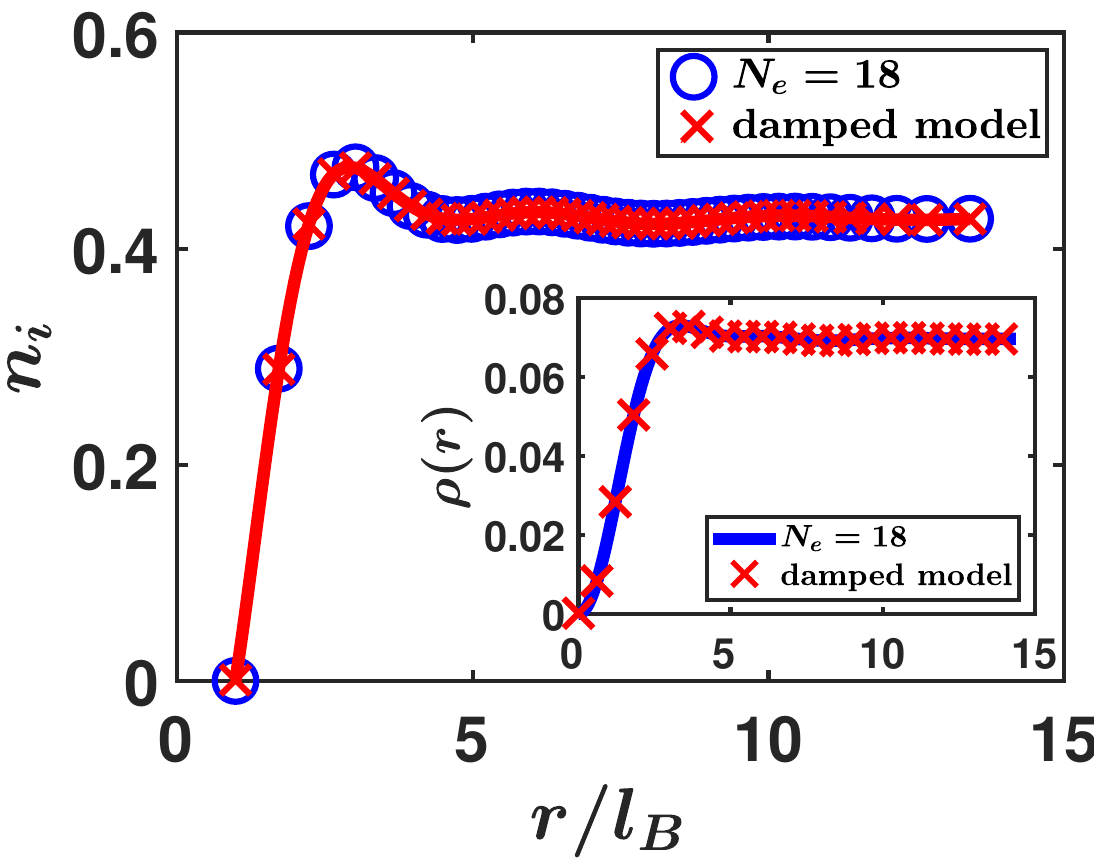}} 
\subfigure[Fibonacci quasihole]{\includegraphics[width=0.48\columnwidth]{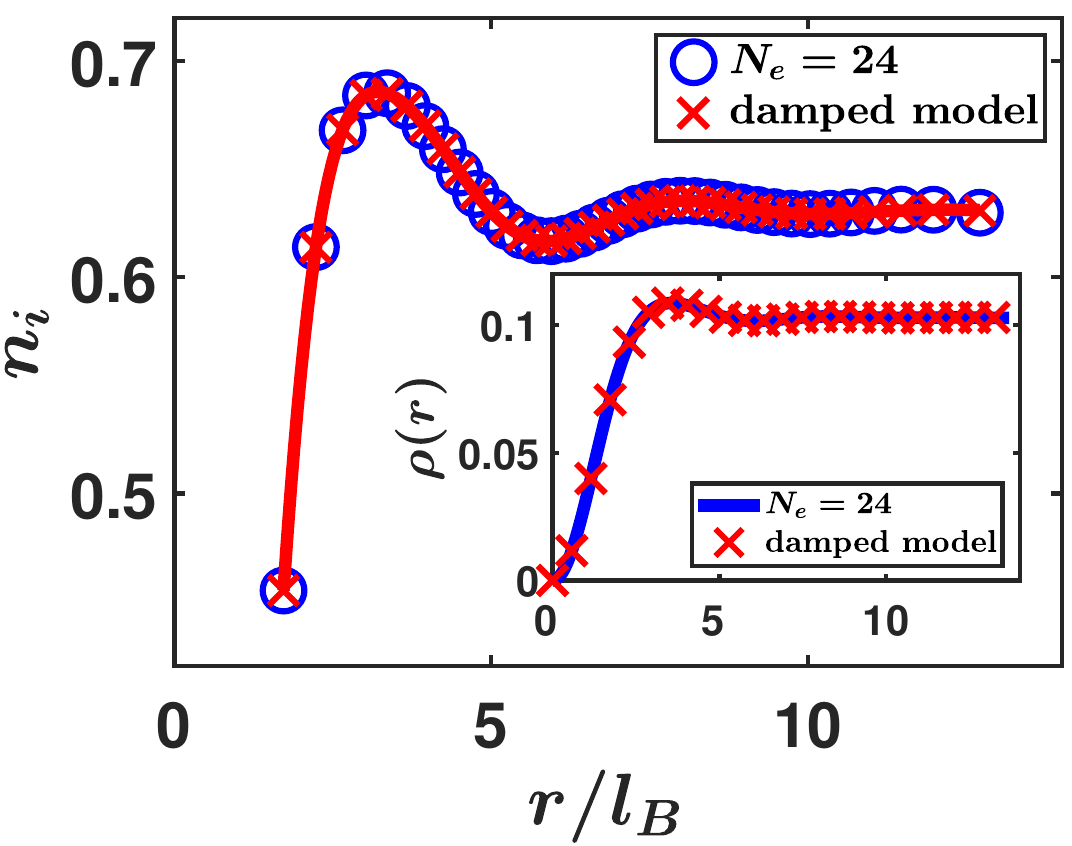}}\subfigure[Jain's quasihole at $\nu=2/5$]{\includegraphics[width=0.48\columnwidth]{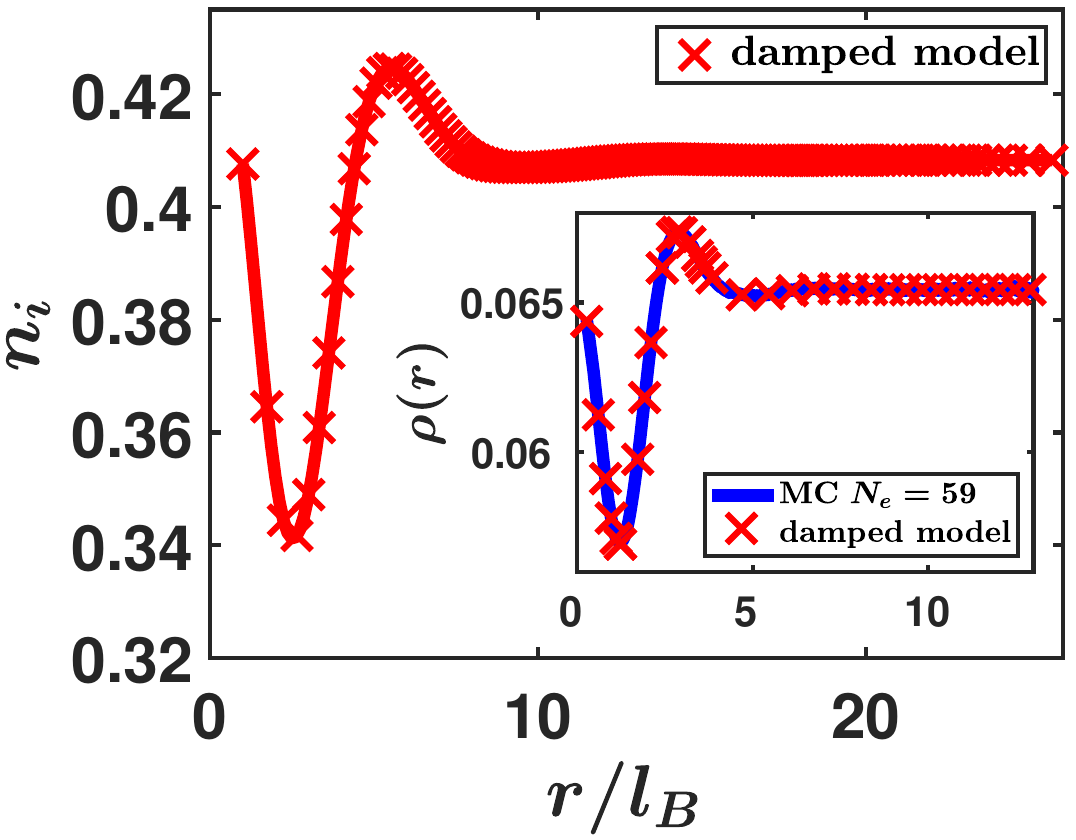}}

\caption{\label{fig:non-Abelian_QH}(a) The occupation numbers of the Moore-Read quasihole for $N_{e}{=}22$ electrons and $N_{\phi}{=}42$ fluxes in the spherical geometry. Blue circles (red crosses) denote the exact (fitted) density generated by Jack polynomials with the root $0110011…0011$. Note that here we only consider the Abelian quasihole consisting of two stacked elementary quasiholes and don’t consider its fractionalization. The inset is the corresponding real-space density. (b) The occupation numbers of the Gaffnian quasihole for $N_{e}{=}18$ electrons and $N_{\phi}{=}42$ fluxes generated by Jack polynomials with the root $01100011…00011$.  (c) The occupation numbers of the Fibonacci quasihole for $N_{e}{=}24$ electrons and $N_{\phi}{=}38$ fluxes generated by Jack polynomials with the root $011100111…00111$. (d) The occupation numbers of the Jain's quasihole for $N_{e}{=}59$ electrons and $N_{\phi}{=}144$ fluxes at $\nu{=}2/5$ generated by the Monte Carlo method.
}
\end{figure}

The generalized damped oscillation model fails to capture the states in higher LLs where the interaction is more long-ranged and the finite-size effects are stronger. Interestingly, the model continues to give a good description of \emph{Abelian quasiholes} obtained from flux insertion (but no fractionalization) for many model FQH fluids [see Fig.~\ref{fig:non-Abelian_QH}], including the Moore-Read~\citep{moore_nonabelions_1991, read_paired_2000}, Gaffnian~\citep{Simon07b}, Fibonacci ~\citep{read_beyond_1999}, and composite fermion states at $\nu{=}2/5$~\citep{Jain89, jain2007composite}. In addition to a very small fitting error, the modeled density profile is much smoother than those obtained from the polynomial expansion method. This is especially the case when the density profile in the real space looks irregular, for example in the case of pair-correlation of the Moore-Read state [see Appendix.~\ref{sec:pair_MooreRead}]. It is important to note that the finite-size scaling of the generalized model is not good suggesting potential over-fitting issues. This technical problem can be overcome by collecting data for larger system sizes by employing the density matrix renormalization group~\citep{zhao_fractional_2011} and Monte Carlo method~\citep{mitra_angular-momentum-state_1993,yang_monte_2023}. That would then allow for large-scale numerical computation of states beyond the simple Laughlin phases.

\section{Quasielectron}
\label{sec:Quasielectron}
\begin{figure}[t]
\subfigure[$V_1$ Quasielectron]{\includegraphics[width=0.48\columnwidth]{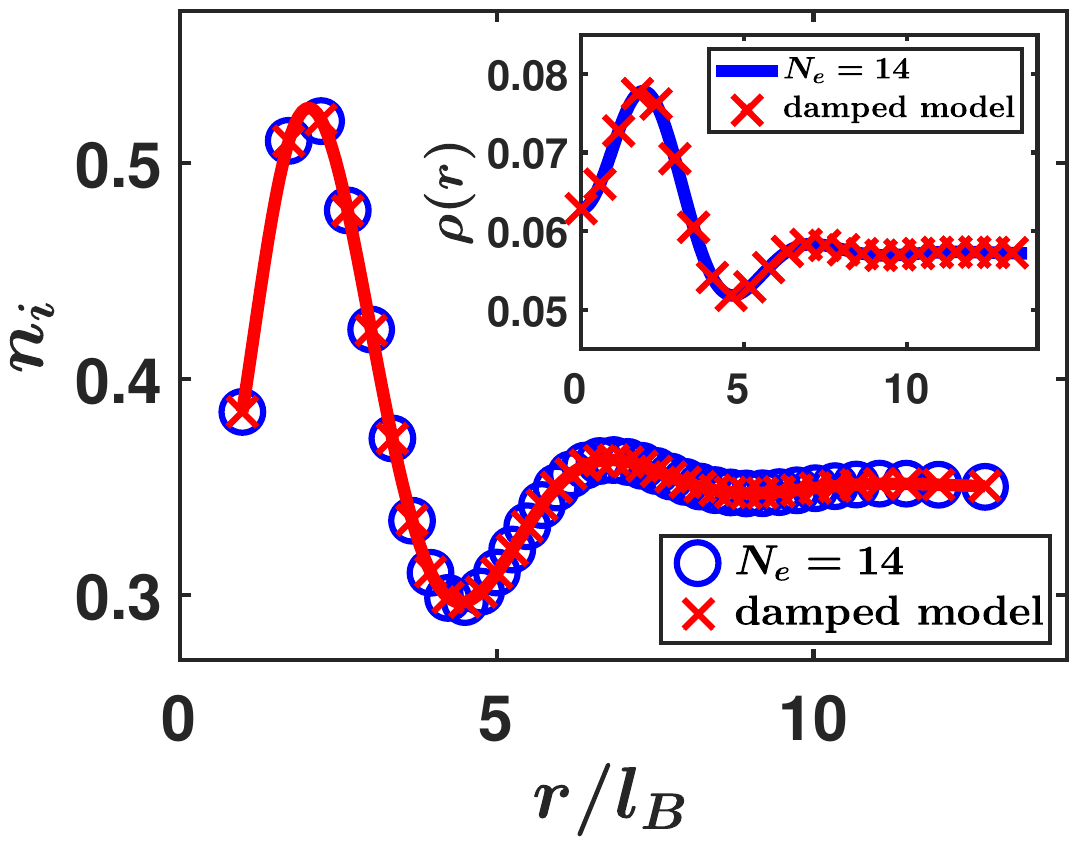}}\subfigure[Coulomb Quasielectron]{\includegraphics[width=0.48\columnwidth]{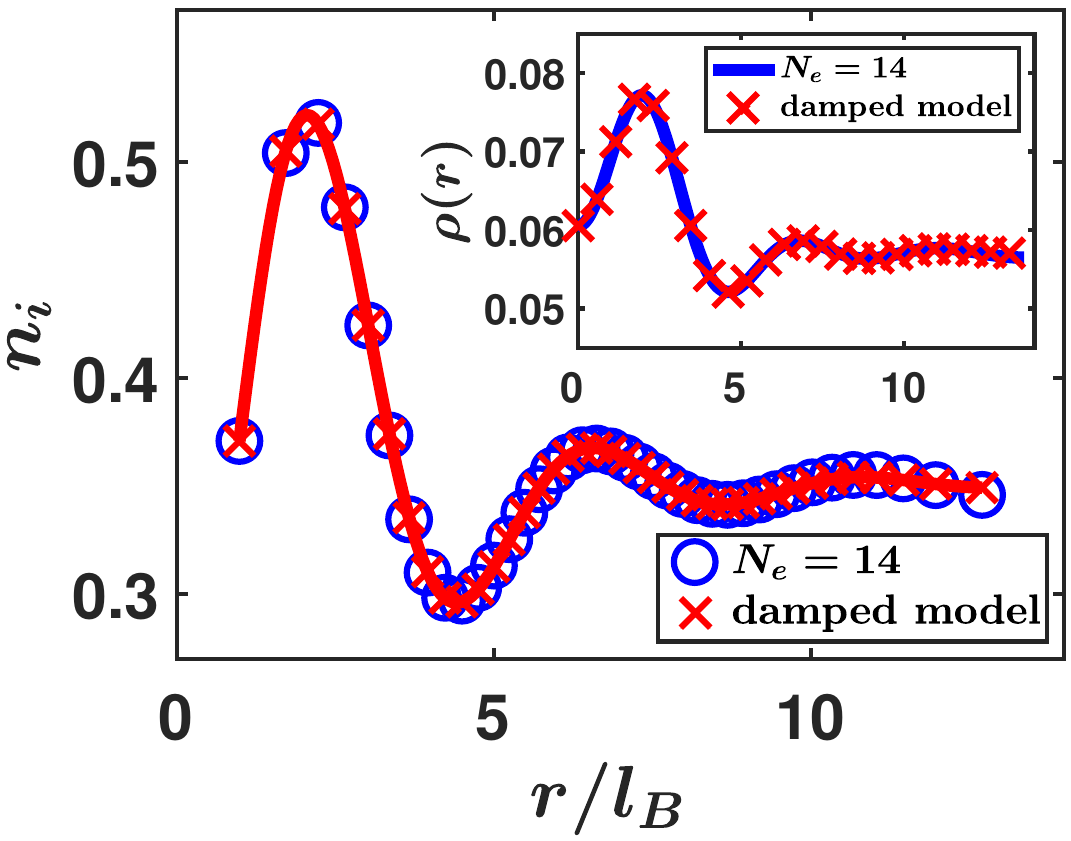}} 
\caption{\label{fig: quasielectron}(a) The occupation numbers of the quasielectron for $N_{e}{=}14$ electrons and $N_{\phi}{=}38$ fluxes in the spherical geometry. Blue circles (red crosses) denote the exact (fitted) density generated by ED with $V_{1}$ interaction. The fitted curve (red line) is the sum of the long-range $0.61\sin[1.40r{-}1.99]\exp({-}0.56r)$ and the short-range ${-}0.73\sin[0.65r{-}1.60]\exp({-}0.96r)$ modes.  The inset is the corresponding real-space density. (b) Similar to (a) except that the interaction used is the Coulomb one. The fitted curve (red line) is the sum of the long-range $0.32\sin[1.47r{-}1.98]\exp({-}0.41r)$ and the short-range ${-}0.24\sin[0.53r{+}2.03]\exp({-}0.64r)$ modes.
}
\end{figure}

The generalized damped oscillator model also describes quasielectron states that result from flux removal in FQH fluids. In contrast to the charge deficiency resulting from a quasihole, a quasielectron at the origin leads to an excess charge. Interestingly, the density oscillations in quasielectron states are very similar to the quasihole ones. In Fig.~\ref{fig: quasielectron}(a) and (b), we plot the density distribution and fitting results for quasielectron states at $\nu{=}1/3$ with $V_{1}$ and Coulomb interactions, respectively. For both states, our generalized model gives a good description. The characteristic lengths of the long-range mode of the Coulomb quasielectron are very close to those of the Coulomb quasihole. This indicates that their density oscillations stem from the same mechanism. 

We note here that in Ref.~\citep{cardoso_boundary_2021}, it is proposed that the density oscillation of the Laughlin edge is due to the formation of the Wigner crystal at the boundary at a low density. However, a quasielectron state (which has an excess charge and higher density than the background fluid) also exhibits density oscillations. Therefore, the crystallization proposal appears unable to explain the density oscillations observed in quasielectron states. These results cast doubt on the crystallization argument for the density oscillation of FQH fluids.

\section{Summary}
\label{sec:Summary}
We show the complex density oscillation of generic Laughlin $n$-quasiholes state in real space can be modeled by a simple damped oscillation in the occupation-number space. Moreover, a generalized damped oscillation model can well-fit the more realistic Coulomb quasihole. The generalized model also applies to many other types of quasiholes. Our work paves the way to reveal the underlying connections between various oscillatory features of FQH fluids and the structure of quasiholes in the occupation-number space. It also provides an avenue to carry out large-scale numerical computations of ground-state variational energies. Moreover, determining the length scales of realistic quasiholes from our model can be useful for designing experimental setups that can measure their fractional charge and statistics.

\begin{acknowledgments}
G. Ji thanks Nicolas Regnault, Ha Quang Trung, Yuzhu Wang, Greg J. Henderson, and Yayun Hu for valuable discussions. Some of the numerical calculations reported in this work have been carried out with the DiagHam package~\cite{diagham} for which we are grateful to its authors. Some of the numerical computations were done on the Nandadevi supercomputer, which is maintained and supported by the Institute of Mathematical Science's High-Performance Computing Center. ACB thanks the Science and Engineering Research Board (SERB) of the Department of Science and Technology (DST) for funding support via the Mathematical Research Impact Centric Support (MATRICS) Grant No. MTR/2023/000002. This work is supported by the National Research Foundation, Singapore under the NRF fellowship award (NRF-NRFF12-2020-005).
\end{acknowledgments}

\appendix

\section{FQH states and Jack polynomials}
\label{sec:Jack}
In this appendix, we show the details of determining the occupation number distribution by Jack polynomials.

Many FQH states including the Laughlin, Moore-Read, Gaffnian, Fibonacci states, and so on can be expressed as Jack polynomials $J_{\bm{\lambda}}^{\alpha}(\{z_i\})$, where $\bm{\lambda}$ is the root of bases, $\alpha$ is a factor determined by the root configuration, and the Gaussian factor $\exp ({-}\sum_{i} r_{i}^2/4)$ has been ignored~\cite{bernevig_model_2008}. We can determine the occupation number distribution of these states by 
\begin{equation}
    n_{i}= \frac{\sum_{\lambda_{s}}  \vert N_{\lambda_{s}} c_{\lambda_{s}}^{\alpha} \vert^{2} \eta_{i}(\lambda_{s})}{\sum_{\lambda_{s}}  \vert N_{\lambda_{s}} c_{\lambda_{s}}^{\alpha} \vert^{2}},
\end{equation}
where $\lambda_{s}$'s are basis configurations squeezed from the root $\bm{\lambda}$, $N_{\lambda_{s}}$ is a geometry-and-occupation-number-dependent normalization factor for each basis denoted by $\lambda_{s}$, $c_{\lambda_{s}}^{\alpha}$ is the Jack coefficient of the basis in $J_{\bm{\lambda}}^{\alpha}(\{z_i \})$, $\eta_{i}(\lambda_{s})$ is the number of the $i^{\text{th}}$ orbital in the basis configuration $\lambda_{s}$. For example, for the Laughlin state at $\nu=1/3$, $\alpha=-2$, $\bm{\lambda}=[100...1001]$, $N_{\lambda_{s}}= \prod_{m_{j}} \sqrt{ 2^{m_{j}} m_{j}!}$ with $m_{j}$ being the angular momentum of occupied orbitals in $\lambda_{s}$ on the disk. The generalization to other states and geometries is straightforward. Since these FQH ground states obey the highest-weight and lowest-weight conditions simultaneously, their density distributions are uniform and $n_{i}{=}\nu$ is a constant. 

In fact, their quasihole excitations induced by flux insertion can also be expressed as Jack polynomials with the same Jack coefficients but different roots. For example, the root for the Laughlin quasihole becomes $\bm{\lambda}{=}[00...0100...1001]$. This leads to a different normalization factor for quasihole states, and their occupation number distributions become
\begin{equation}
    n^{\text{QH}}_{i}= \frac{\sum_{\lambda_{s}}  \vert N^{\text{QH}}_{\lambda_{s}} c_{\lambda_{s}}^{\alpha} \vert^{2} \eta_{i}(\lambda_{s})}{\sum_{\lambda_{s}}  \vert N^{\text{QH}}_{\lambda_{s}} c_{\lambda_{s}}^{\alpha} \vert^{2}},
\end{equation}
where the normalization factor of the FQH $n$-quasihole state on the disk is related to that of the ground state by $ N^{\text{QH}}_{\lambda_{s}} {=} N_{\lambda_{s}} \prod_{m_{j}} \prod_{l{=}1}^{n} \sqrt{ m_{j}{+}l}$. This change results in the density oscillation of FQH quasihole states. Even though the change for each basis $\lambda_{s}$ can be traced as shown above, the large number of basis (e.g., more than $2{\times}10^{10}$ basis states for the Laughlin quasihole with $N_{e}{=}17$) prevents us from deriving an analytical expression for its density distribution. Therefore, we have to calculate it numerically and try to capture its physics by finding the best fitting for its distribution with as few fitting parameters as possible as done in the main text.

\section{1/3 Laughlin quasihole}
\label{sec:1over3_QH}
In this Appendix and following Appendices, we show the detailed numerical data of fitting parameters for the model $n$-Laughlin quasiholes at $\nu{=}1/m$, the Coulomb quasihole at $\nu{=}1/3$ and the pair-correlation function of Moore-Read state at $\nu{=}1/2$.

\begin{figure}[htpb]
\includegraphics[width=0.48\columnwidth]{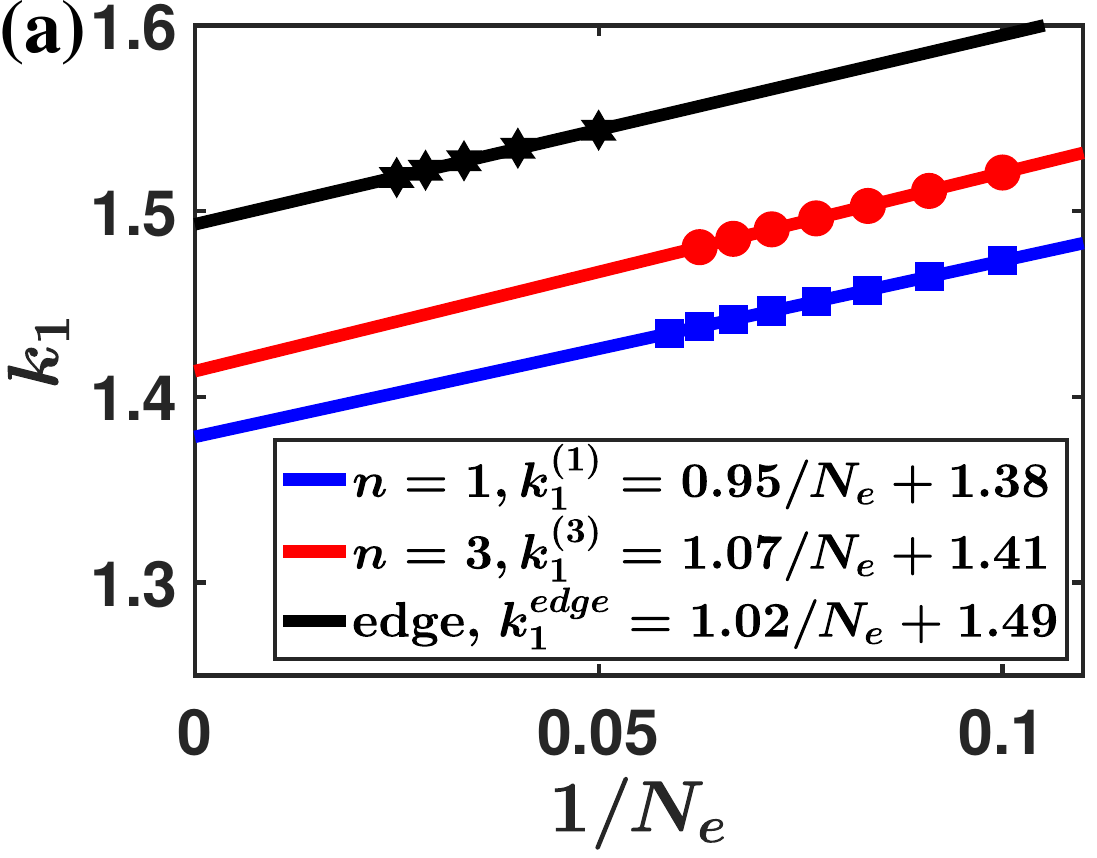}\hfill{}\includegraphics[width=0.48\columnwidth]{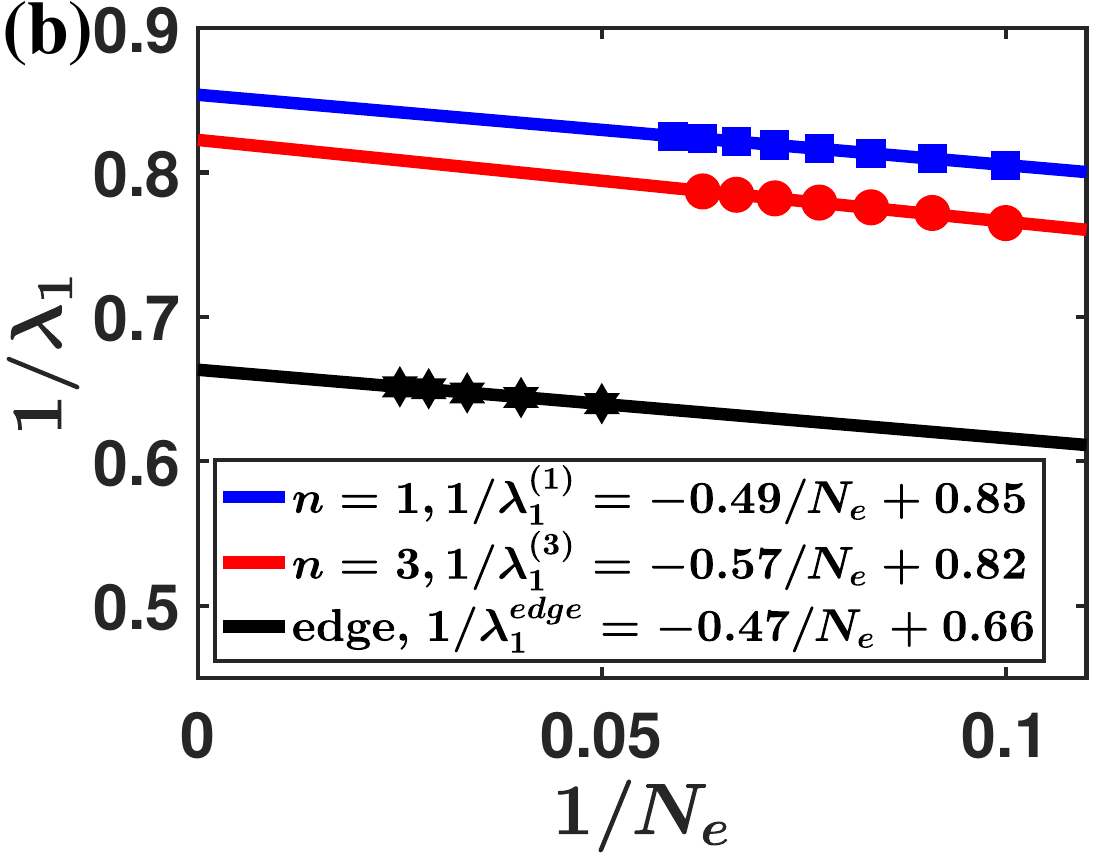}
\caption{\label{fig: finite_size_sclaing_m_3}
(a) Finite-size scaling of the wave vector $k_{1}$ for the single Laughlin quasihole (blue squares), $3$-Laughlin quasiholes (red circles), and the Laughlin edge (black hexagrams) at $\nu{=}1/3$. The lines are linear fits as a function of $1/N_{e}$. (b) Finite-size scaling of the inverse of the decay length $1/\lambda_{1}$. 
}
\end{figure}

We obtain the occupation number $n_{i}$ of a single Laughlin quasihole for $m{=}3$ and $N_{e}{=}10{-}17$ using Jack polynomials~\cite{bernevig_model_2008}. The parameters of the damped oscillatory model $\delta n_{i}{=}A_{1}\sin[k_{1}(x_{i}{-}x_{1})]\exp\left({-}x_{i}/\lambda_{1}\right)$ for the occupation-number $\delta n_{i}{=}n_{i}{-}\bar{n}$ are shown in Table ~\ref{table:fitting_parameters_m_3_n_1}. The finite size scaling of the oscillation wave number and the decay length is shown in Fig.~\ref{fig: finite_size_sclaing_m_3}. They have an almost perfect linear scaling in $1/N_{e}$ allowing us to do the thermodynamic extrapolation reliably. Once the $n_{i}$ are determined, the real-space density of Laughlin quasihole in the thermodynamic limit is given by
\begin{align}
    \rho (r) &=\sum_{i=0}^{\infty} n_{i} \rho_{i}(r), \\
    \rho_{i}(r) &=\frac{1}{2\pi 2^{i} i!}r^{2i}\exp \left[ -\frac{r^2}{2} \right].
\end{align}
To check the accuracy of the model, we compare the density profile of the quasihole in the thermodynamic limit produced by our method and the polynomial expansion method proposed in Ref.~\citep{fulsebakke_parametrization_2023}. The two methods deviate very slightly from each other (the root-mean-square difference is $2{\times}10^{-5}$) producing nearly coincident curves as shown in Fig.~\ref{fig: single_quasihole}(a). This implies that, even though our model is very simple with only two free parameters, our result is as accurate as the polynomial expansion method which uses several tens of parameters.

\begin{table*}[htpb!]
\centering
\renewcommand{\arraystretch}{1.2}
\begin{tabular}{ | m{0.8cm}<{\centering} |m{1.8cm}<{\centering} |m{1.8cm}<{\centering} |m{1.8cm}<{\centering} |m{1.8cm}<{\centering} |m{1.8cm}<{\centering}  |}	
\hline 
$N_{e}$ & $A_{1}$ & $k_{1}$ & $k_{1}*x_{1}$ & $1/\lambda_{1}$ & $R^{2}$ \\ \hline
10	&	0.6757	&	1.4734	&	3.3342	&	0.8051	&	0.999883	\\ \hline
11	&	0.6774	&	1.4646	&	3.3305	&	0.8097	&	0.999893 \\ \hline
12	&	0.6787	&	1.4574	&	3.3277	&	0.8133	&	0.999904 \\ \hline
13	&	0.6799	&	1.4514	&	3.3253	&	0.8164	&	0.999910 \\ \hline
14	&	0.6809	&	1.4462	&	3.3231	&	0.8191	&	0.999915 \\ \hline
15	&	0.6819	&	1.4417	&	3.3211	&	0.8214	&	0.999918 \\ \hline
16	&	0.6827	&	1.4377	&	3.3194	&	0.8234	&	0.999921 \\ \hline
17	&	0.6835	&	1.4343	&	3.3178	&	0.8252	&	0.999923 \\ \hline
$\infty$	&	0.6944	&	1.3785 &	3.2947	&	0.8538	&	0.999982 \\ \hline
\end{tabular} 
\caption{\label{table:fitting_parameters_m_3_n_1} Parameters of the damped oscillator model for the density of a single Laughlin quasihole at $\nu{=}1/3$ for $N_{e}$ electrons in the spherical geometry. The statistical measure $R^{2}$ is the coefficient of determination for how well the modeled occupation predicts the exact occupation numbers. The final row is the thermodynamic extrapolation result as a function of $1/N_{e}$. }
\end{table*}

\begin{table*}[htpb!]
\centering
\renewcommand{\arraystretch}{1.2}
\begin{tabular}{ | m{0.8cm}<{\centering} |m{1.8cm}<{\centering} |m{1.8cm}<{\centering} |m{1.8cm}<{\centering} |m{1.8cm}<{\centering} |m{1.8cm}<{\centering}  |}	
\hline 
$N_{e}$ & $A_{1}$ & $k_{1}$ & $k_{1}*x_{1}$ & 1/$\lambda_{1}$ & $R^{2}$ \\ \hline
10	&	2.1345	&	1.5208	&	4.8670	&	0.7670	&	0.999885\\ \hline
11	&	2.1446	&	1.5109	&	4.8520	&	0.7720	&	0.999886\\ \hline
12	&	2.1514	&	1.5028	&	4.8402	&	0.7759	&	0.999885\\ \hline
13	&	2.1571	&	1.4961	&	4.8304	&	0.7792	&	0.999887\\ \hline
14	&	2.1625	&	1.4902	&	4.8219	&	0.7821	&	0.999888\\ \hline
15	&	2.1676	&	1.4851	&	4.8143	&	0.7846	&	0.999890\\ \hline
16	&	2.1724	&	1.4806	&	4.8076	&	0.7869	&	0.999891\\ \hline
$\infty$ &	2.2335	&	1.4138	&	4.7094	&	0.8198	&	0.999900\\ \hline
\end{tabular} 
\caption{\label{table:fitting_parameters_m_3_n_3} Similar to Table ~\ref{table:fitting_parameters_m_3_n_1} except the parameters are for $3{-}$Laughlin quasiholes at $\nu{=}1/3$.}
\end{table*}

\begin{table*}[htpb!]
\centering
\renewcommand{\arraystretch}{1.2}
\begin{tabular}{ | m{0.8cm}<{\centering} |m{1.8cm}<{\centering} |m{1.8cm}<{\centering} |m{1.8cm}<{\centering} |m{1.8cm}<{\centering} |m{1.8cm}<{\centering}  |}	
\hline 
$N_{e}$ & $A_{1}$ & $k_{1}$ & $k_{1}*x_{1}$ & 1/$\lambda_{1}$ & $R^{2}$ \\ \hline
20	&	5.2491e-04	&	1.5436	&	12.5625	&	0.6400	&	0.999889\\ \hline
25	&	2.1314e-04	&	1.5339	&	14.4496	&	0.6442	&	0.999894\\ \hline
30	&	9.4669e-05	&	1.5271	&	16.1576	&	0.6474	&	0.999899\\ \hline
35	&	4.4788e-05	&	1.5220	&	17.7294	&	0.6503	&	0.999904\\ \hline
40	&	2.2664e-05	&	1.5181	&	19.1959	&	0.6516	&	0.999905\\ \hline
$\infty$ &	0	&	1.4929	&	$\infty$	&	0.6635	&	0.999922\\ \hline
\end{tabular} 
\caption{\label{table:fitting_parameters_edge_m_3} Parameters for the Laughlin edge with $N_{e}$ electrons at $\nu{=}1/3$. $R^{2}$ is the R-squared value for the real-space density obtained by the Monte Carlo method and the modeled real-space density in the planar disk geometry. In the thermodynamic limit, $A_{1}{=}0$ and $x_{1}{=}\infty$ because the edge is located at the position $r{=}\infty$. Values in the last row are the thermodynamic extrapolation results obtained from a linear fit of the parameters as a function of $1/N_{e}$.}
\end{table*}

\begin{table*}[htpb!]
\centering
\renewcommand{\arraystretch}{1.2}
\begin{tabular}{ | m{0.8cm}<{\centering} |m{1.8cm}<{\centering} |m{1.8cm}<{\centering} |m{1.8cm}<{\centering} |m{1.8cm}<{\centering} |m{1.8cm}<{\centering}  |}
\hline 
$N_{e}$ & $A_{1}$ & $k_{1}$ & $k_{1}*x_{1}$ & 1/$\lambda_{1}$ & $R^{2}$ \\ \hline
15 &	4.6540	&	1.1986	&	5.1067	&	0.4515	&	0.999970\\ \hline
20 &	4.7661	&	1.1839	&	5.0775	&	0.4595	&	0.999974\\ \hline
25 &	4.8301	&	1.1755	&	5.0613	&	0.4640	&	0.999976\\ \hline
30 &	4.8715	&	1.1701	&	5.0510	&	0.4669	&	0.999977\\ \hline
50 &	4.9478	&	1.1596	&	5.0317	&	0.4722	&	0.999979\\ \hline
$\infty$ &	5.0791	&	1.1425	&	4.9980	&	0.4815	&	0.999984\\ \hline
\end{tabular} 
\caption{\label{table:fitting_parameters_m_5} Similar to Table ~\ref{table:fitting_parameters_edge_m_3} except the parameters are for the ground state pair-correlation function at $\nu{=}1/5$.}
\end{table*}

\section{1/3 Laughlin state pair-correlation function}
\label{sec:1over3_pair}
We repeat the single quasihole analysis for $3$-Laughlin quasiholes at $\nu{=}1/3$. The parameters are shown in Table ~\ref{table:fitting_parameters_m_3_n_3} and the finite size scaling is shown in Fig.~\ref{fig: finite_size_sclaing_m_3}. The ground state pair-correlation function $g(r)$ in the thermodynamic limit can be obtained by dividing the density $\rho(r)$ of $3$-Laughlin quasiholes by the average density $\bar{\rho}{=}1/(2 \pi m)$. Then, one can use the modeled $g(r)$ to calculate the per-particle variational energy of the Laughlin state for various interactions $V(r)$ through $V{=}(\bar{\rho}/2) \int d^{2} \, \bm{r} V(r)[g(r){-}1]$. For the Coulomb interaction $V(r){=}1/r$, the calculated energy $V$ with our modeled $g(r)$ is ${-}0.4096$, which is very close to the extrapolated result ${-}0.4098$ obtained from the exact diagonalization calculation~\citep{balram_fractional_2020}.

\section{1/3 Laughlin edge}
\label{sec:1over3_edge}
\begin{figure}[htpb!]
\begin{center}
\includegraphics[width=0.9\columnwidth]{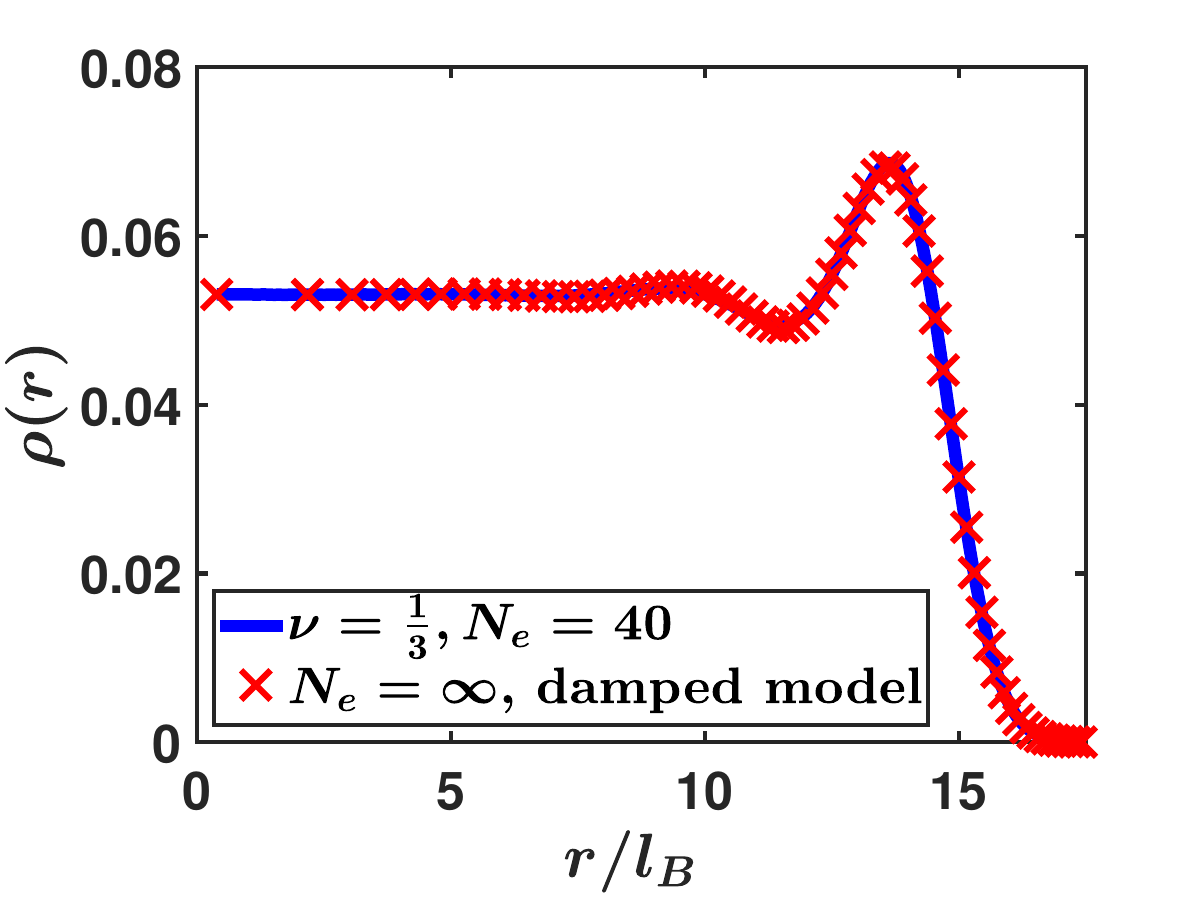}
\end{center}
\caption{The density distribution of $1/3$ Laughlin state for $N_{e}=40$ in the disk (blue line). The red crosses denote the fitting using our damped model.}
\label{fig: plot_edge_fitting}
\end{figure}

We create a Laughlin edge by placing the Laughlin state on the disk geometry. Since the edge density fluctuation is large, we need a large system size (beyond the ones accessible to exact diagonalization) to reliably study it. Therefore, we determine the edge density using the Monte Carlo method. Then, we obtain the parameters of the damped model by fitting them to the real-space density. As shown in Fig.~\ref{fig: plot_edge_fitting}, the real-space density profile of the Laughlin edge can also be well-represented with the damped model in the occupation-number space. The parameters for different sizes are shown in Table ~\ref{table:fitting_parameters_edge_m_3} and the finite size scaling is shown in Fig.~\ref{fig: finite_size_sclaing_m_3}. 

\section{1/5 Laughlin state pair-correlation function}
\label{sec:1over5_pair}
\begin{figure}[htpb!]
\begin{center}
\includegraphics[width=0.9\columnwidth]{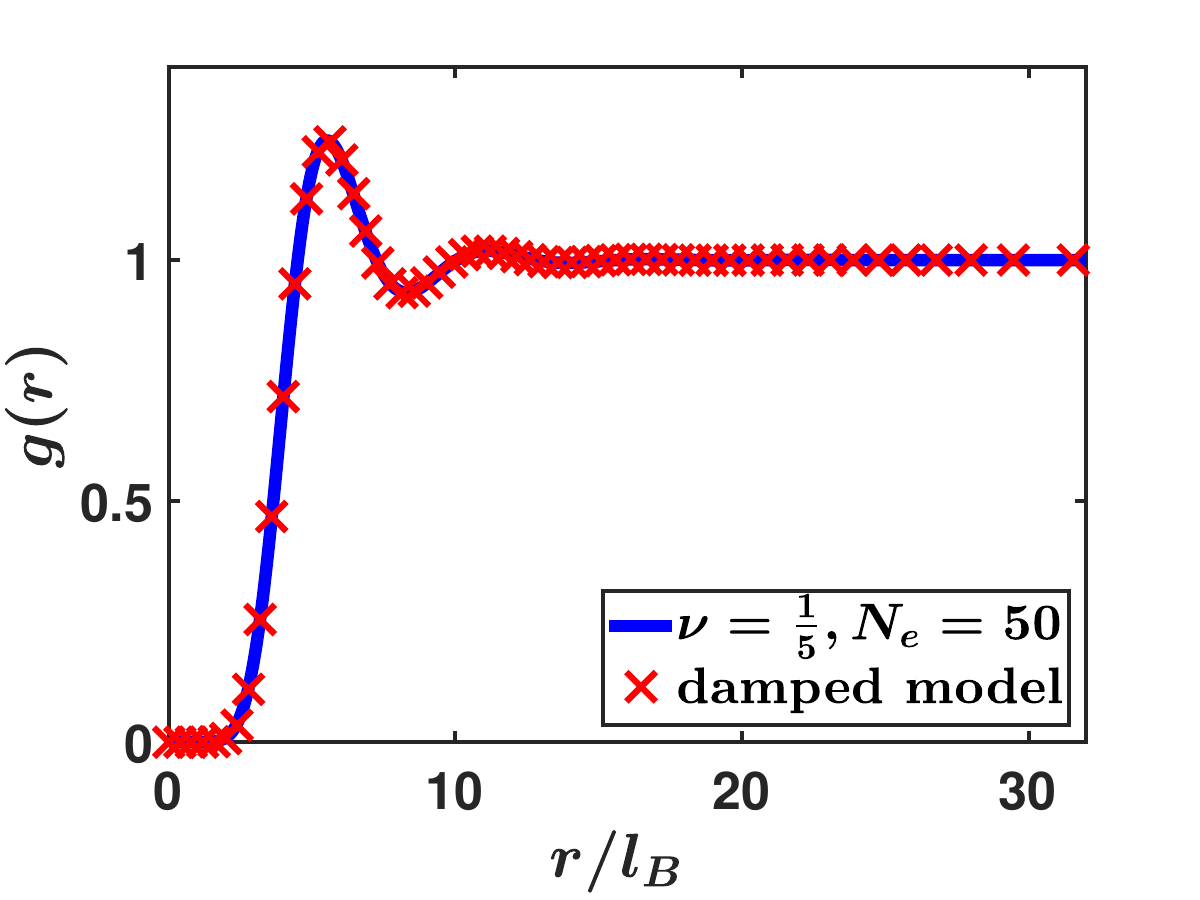}
\end{center}
\caption{The ground state pair-correlation function for the $1/5$ Laughlin state for $N_{e}{=}50$ electrons in the spherical geometry (blue line). The red crosses denote the results obtained using our damped model.}
\label{fig: pair_correlation_function_m_5}
\end{figure}

We also study the ground state pair-correlation function for $\nu{=}1/5$ Laughlin state using the Monte Carlo method and determine the parameters of the damped model by fitting the model to the real-space profile. As shown in Fig.~\ref{fig: pair_correlation_function_m_5}, the pair-correlation function for the 1/5 Laughlin state can also be well-represented by our damped model. The parameters for different sizes and their thermodynamic extrapolation are shown in Table ~\ref{table:fitting_parameters_m_5}. Then, we use the modeled $g(r)$ to calculate the per-particle variational energy of the Laughlin state in terms of the Coulomb interaction $V(r){=}1/r$, and the result is ${-}0.3278$, which is very close to the extrapolated result ${-}0.3275(1)$ obtained from exact diagonalization~\citep{dora2023competition}.

\begin{table*}[htpb!]
\centering
\renewcommand{\arraystretch}{1.2}
\begin{tabular}{ | m{0.8cm}<{\centering} |m{1.5cm}<{\centering} |m{1.5cm}<{\centering} |m{1.5cm}<{\centering} |m{1.5cm}<{\centering} |m{1.5cm}<{\centering}  |m{1.5cm}<{\centering}  |m{1.5cm}<{\centering}  |m{1.5cm}<{\centering}  |m{1.5cm}<{\centering}  |}	
\hline 
$N_{e}$ & $A_{1}$ & $k_{1}$ & $k_{1}*x_{1}$ & $1/\lambda_{1}$ & $A_{2}$ & $k_{2}$ & $k_{2}*x_{2}$ & $1/\lambda_{2}$ & $R^{2}$ \\ \hline
10	&	0.2039	&	1.4965	&	2.7515	&	0.3778	&	0.5390	&	0.6411	&	2.0603	&	0.9143	&	0.999694\\ \hline
11	&	0.2283	&	1.4794	&	2.7356	&	0.4078	&	0.4948	&	0.6186	&	2.0792	&	0.8994	&	0.999920\\ \hline
12	&	0.2317	&	1.5039	&	2.8840	&	0.4202	&	0.5449	&	0.4572	&	1.5763	&	0.9318	&	0.999933\\ \hline
13	&	0.2232	&	1.4982	&	2.8624	&	0.4129	&	0.5375	&	0.5095	&	1.7350	&	0.9445	&	0.999845\\ \hline
14	&	0.2262	&	1.4847	&	2.8172	&	0.4160	&	0.5115	&	0.5425	&	1.8643	&	0.9318	&	0.999861\\ \hline
\end{tabular} 
\caption{\label{table:fitting_parameters_coulomb_qh} Parameters of the double-damped model for the Coulomb quasihole with $N_{e}$ electrons at $\nu{=}1/3$.}
\end{table*}

\begin{table*}[htpb!]
\centering
\renewcommand{\arraystretch}{1.2}
\begin{tabular}{ | m{0.8cm}<{\centering} |m{1.5cm}<{\centering} |m{1.5cm}<{\centering} |m{1.5cm}<{\centering} |m{1.5cm}<{\centering} |m{1.5cm}<{\centering}  |m{1.5cm}<{\centering}  |m{1.5cm}<{\centering}  |m{1.5cm}<{\centering}  |m{1.5cm}<{\centering}  |}	
\hline 
$N_{e}$ & $A_{1}$ & $k_{1}$ & $k_{1}*x_{1}$ & $1/\lambda_{1}$ & $A_{2}$ & $k_{2}$ & $k_{2}*x_{2}$ & $1/\lambda_{2}$ & $R^{2}$ \\ \hline
15	&	0.9540	&	1.5628	&	5.7802	&	0.5936	&	7.3000	&	1.4715	&	3.4053	&	1.5951	&	0.999434\\ \hline
17	&	0.9761	&	1.5274	&	5.6332	&	0.6076	&	7.5191	&	1.4616	&	3.3479	&	1.6510	&	0.999540\\ \hline
19	&	1.0409	&	1.5138	&	5.5905	&	0.6273	&	6.9436	&	1.4817	&	3.3604	&	1.6286	&	0.999566\\ \hline
21	&	1.0593	&	1.5009	&	5.5481	&	0.6354	&	6.6887	&	1.4837	&	3.3545	&	1.6288	&	0.999598\\ \hline
$\infty$		&	1.3437	&	1.3457	&	4.9646	&	0.7451	&	5.0878	&	1.5220	&	3.2328	&	1.7069	&	\\ \hline
\end{tabular} 
\caption{\label{table:fitting_parameters_MooreRead} Parameters of the double-damped model for the Moore-Read hole density proportional to the ground state pair-correlation function. The last row is the thermodynamic extrapolation result as a function of $1/N_{e}$. }
\end{table*}

\section{General Laughlin cases}
\label{sec:general_laughlin}
\begin{figure}[htpb!]
\includegraphics[width=0.48\columnwidth]{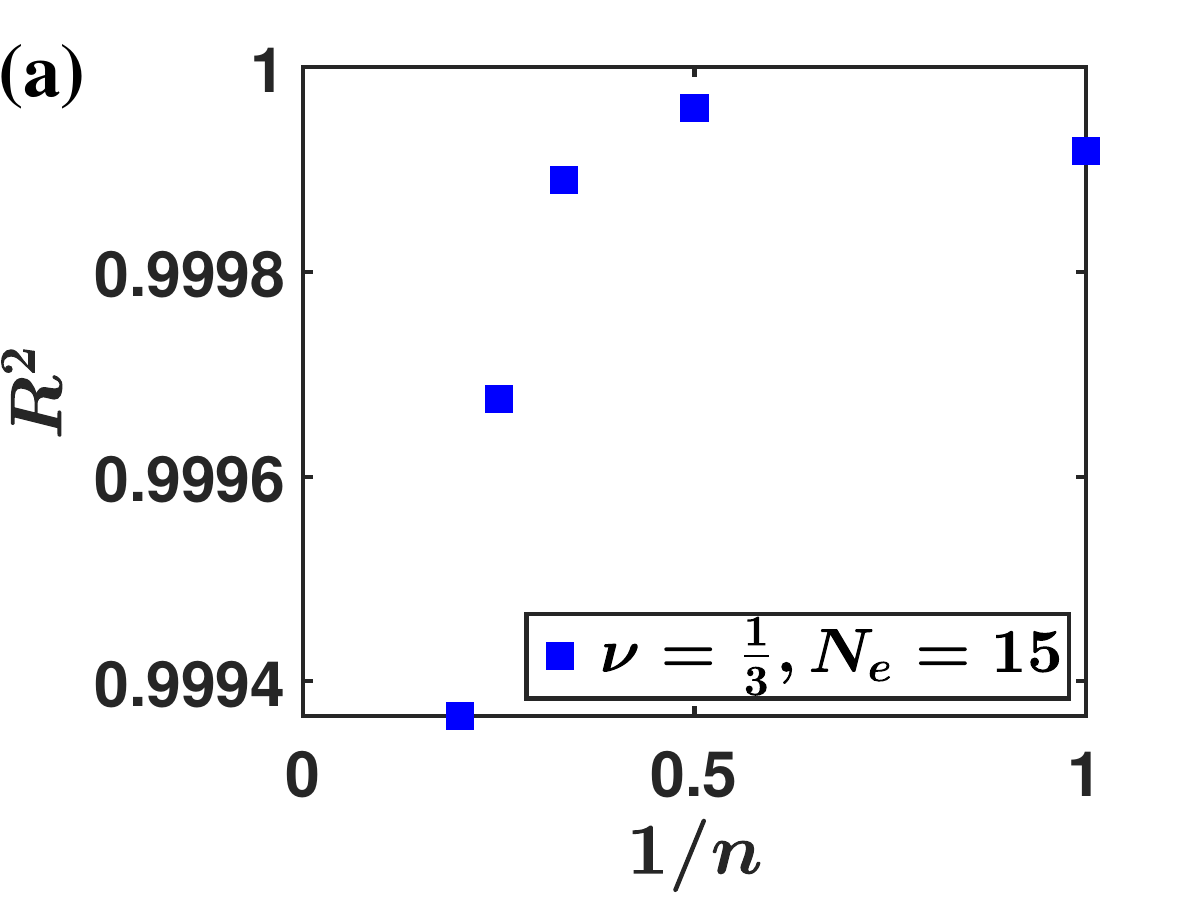}\hfill{}\includegraphics[width=0.48\columnwidth]{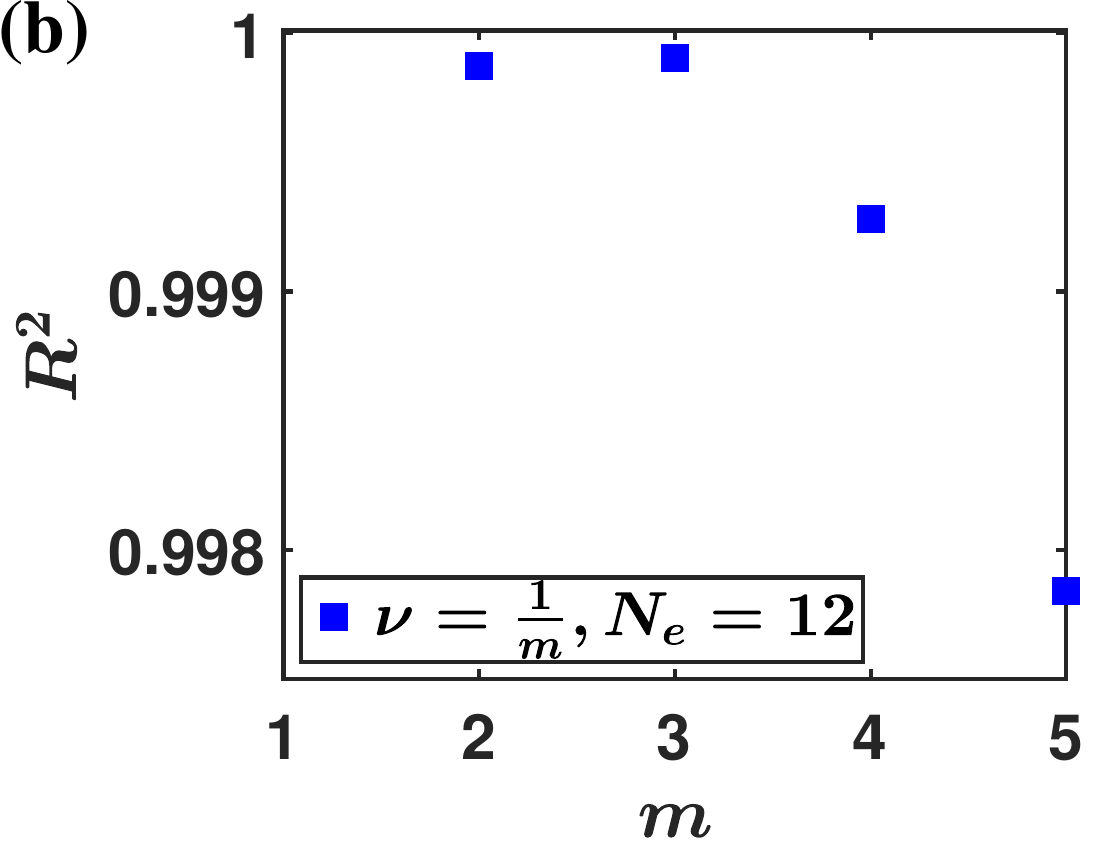}
\caption{\label{fig: R2_analysis}
(a) $R^{2}$ value for Laughlin $n$-quasihole state at $\nu=1/3$ and $N_{e}=15$. The statistical measure $R^{2}$ is the coefficient of determination for how well the modeled occupation predicts the exact occupation numbers obtained by Jack polynomial. (b) $R^{2}$ value for Laughlin single quasiholes at $\nu=1/m$ and $N_{e}=12$. 
}
\end{figure}

By using the Jack polynomial method or Monte Carlo method, we can determine the occupation-number density or real-space density of Laughlin quasihole for various $m$ and $n$. Then, we can determine their characteristic lengths as done above and the results are summarized in Fig.~\ref{fig: characteric_length_nstacked}. Based on the numerical results for various Laughlin quasiholes, we find the damped oscillation model can always reasonably capture the density-distribution of quasiholes, but its numerical accuracy ($R^{2}$ value) decreases increasing $m$ and $n$ as shown in Fig.~\ref{fig: R2_analysis}. For large values of $m$ and $n$, there are stronger finite-size effects which can potentially explain the lower numerical accuracy of the damped model. For these large values of $m$ and $n$, it is also possible that our model is unable to capture the physical features of the states accurately. When the density fluctuation is very strong, especially for large $m$, it is better to determine the characteristic lengths of the oscillation tail by starting from a finite $r$, e.g., the first zero point of $\delta \rho (r){=}0$. 

\section{Coulomb quasihole}
\label{sec:Coulomb_quasihole}
In addition to the single-mode model of the Laughlin quasihole, another mode with four more fitting parameters is introduced for the Coulomb quasihole. The fitting parameters are shown in Table ~\ref{table:fitting_parameters_coulomb_qh}. Note that numerical results show that there exist multiple sets of parameters that fit the density distribution very well. Therefore, it is possible that overfitting could be an issue for this model. We also find strong finite-size-scaling effects that hinder an accurate extrapolation to the thermodynamic limit. 

\begin{figure}[htpb!]
\begin{center}
\includegraphics[width=0.9\columnwidth]{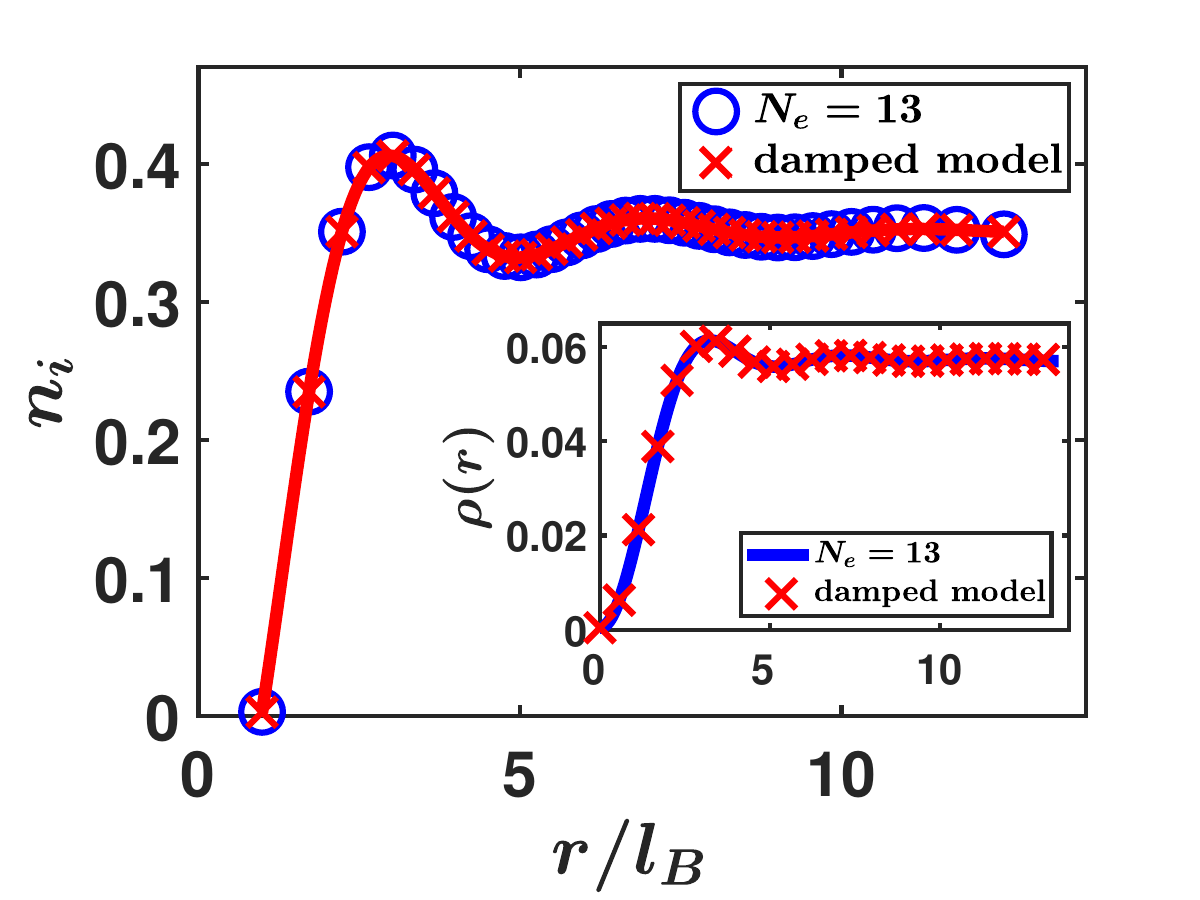}
\end{center}
\caption{The occupation numbers of the quasihole for $N_{e}{=}13$ electrons and $N_{\phi}{=}37$ fluxes in the spherical geometry. Blue circles (red crosses) denote the exact (fitted) density generated by ED with Yukawa interaction $V(r){=}e^{{-}r}/r$. The inset is the corresponding real-space density.}
\label{fig: Yukawa}
\end{figure}

For the correction to the Coulomb potential, we consider the Yukawa interaction $V(r){=}e^{{-}\lambda r}/r$. The overlap of the quasihole state with finite decay length $\lambda{=}l_{B}$ and the bare Coulomb quasihole is $0.9910$. Their density profiles are very close to each other. The double-damped model also works in this case as shown in Fig.~\ref{fig: Yukawa}. If we further increase $\lambda$, the quasihole state will approach the Laughlin quasihole gradually, in which case our model also works. Thus, we expect the double-damped model also works in the case of Yukawa interaction in the LLL.

\section{pair-correlation function of Moore-Read state}
\label{sec:pair_MooreRead}
\begin{figure}[htpb!]
\includegraphics[width=0.48\columnwidth]{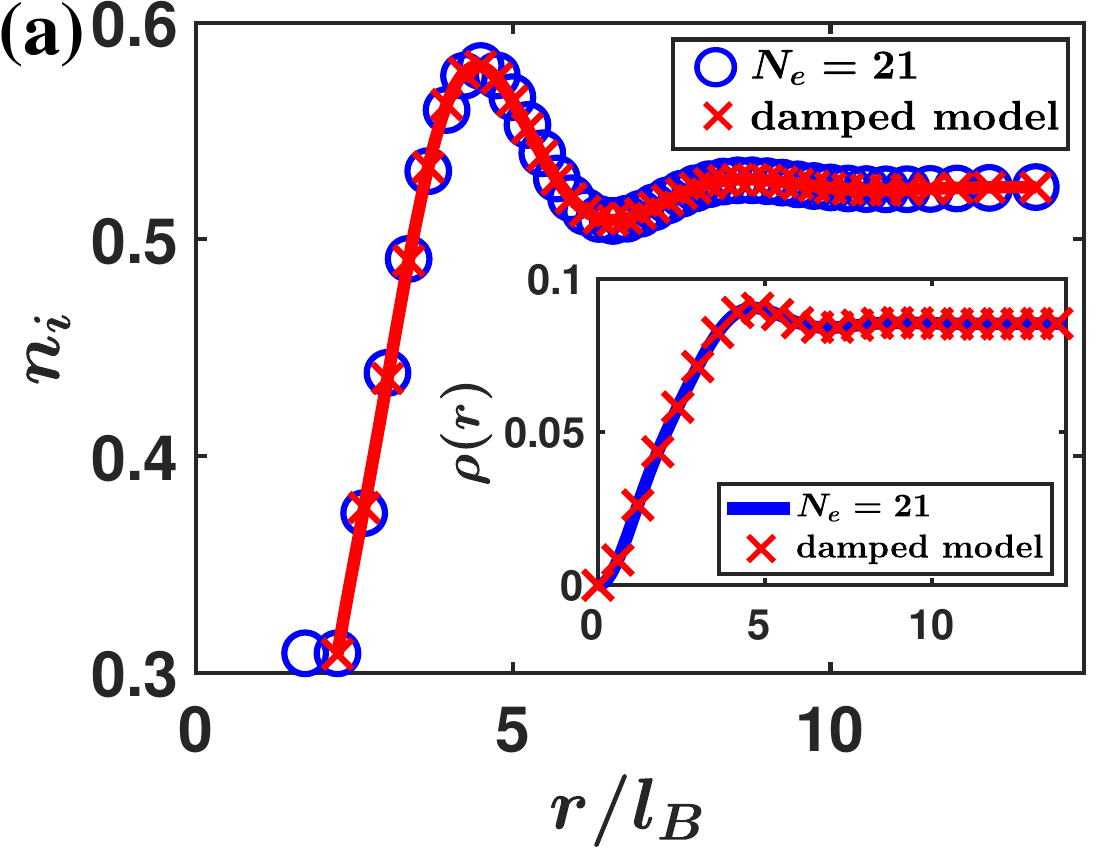}\hfill{}\includegraphics[width=0.48\columnwidth]{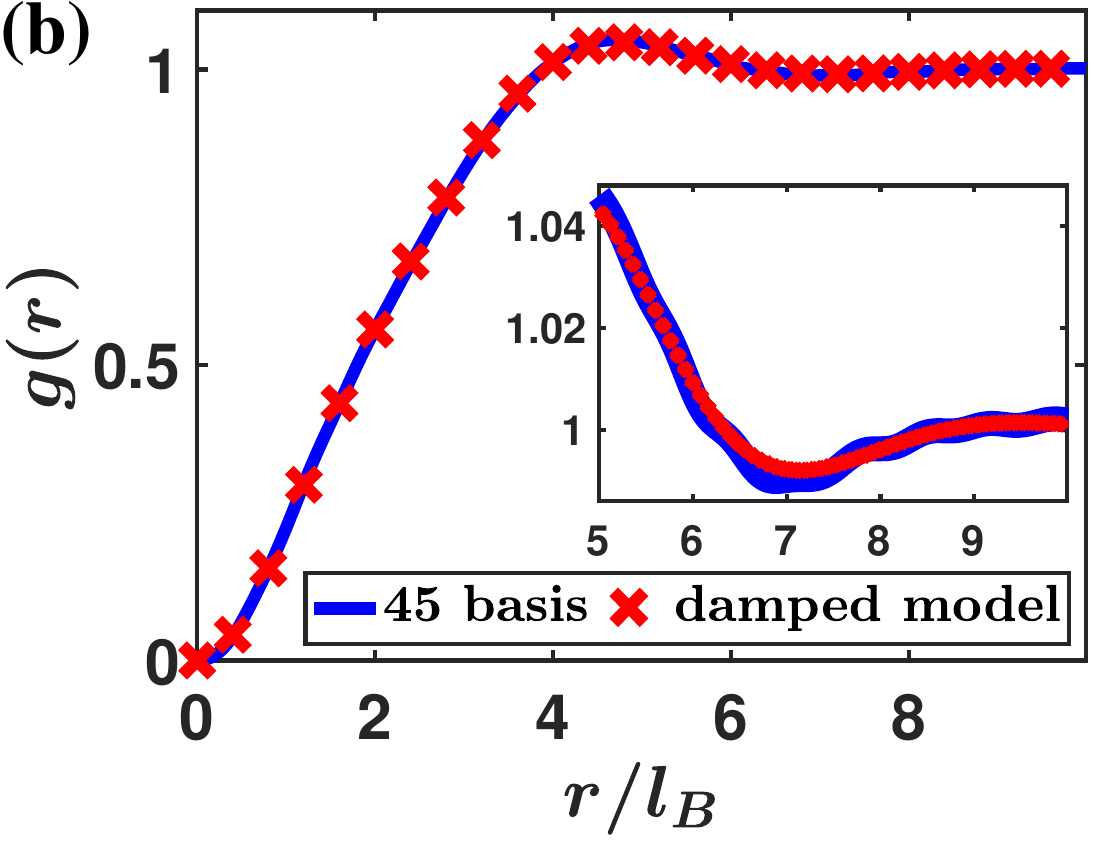}
\caption{\label{fig:fitting_MooreRead}(a) The occupation numbers of the Moore-Read hole state, the density of which is proportional to the Moore-Read ground state pair-correlation function, for $N_{e}{=}21$ electrons and $N_{\phi}{=}41$ fluxes in the spherical geometry. Blue circles (red crosses) denote the exact (fitted) density generated by Jack polynomials with the root $010011...0011$. The inset is the corresponding real-space density. (b) The Moore-Read ground state pair-correlation function in the thermodynamic limit. The blue line and red crosses denote the results derived using the polynomial expansion method with $45$ basis~\citep{fulsebakke_parametrization_2023} and our double-damped model, respectively. The inset is the zoomed view of the oscillation tail.
}
\label{fig: pair_correlation_function_MooreRead}
\end{figure}

As an example, we show how to model the pair-correlation function of the Moore-Read ground state in the thermodynamic limit from our double-damped model. To reduce the finite-size-scaling effects, we first determine the parameters of the long-range mode by fitting with the occupation numbers at a large $r$ (here our fitting starts from $n_{8}$), and then we determine the parameters of the short-range mode using all occupation numbers except the first two special orbitals. In this way, the finite-size effect is greatly reduced and meanwhile, the accuracy is still good as shown in Table ~\ref{table:fitting_parameters_MooreRead} and Fig.~\ref{fig:fitting_MooreRead}(a). The result obtained from an extrapolation to the thermodynamic limit is shown in Fig.~\ref{fig:fitting_MooreRead}(b). For comparison, we also plot the corresponding result obtained from the polynomial expansion method~\citep{fulsebakke_parametrization_2023}. The two methods produce very similar results with a root-mean-square difference of $2{\times}10^{-3}$). Nevertheless, the density profile obtained from our model is smoother than that obtained using the polynomial expansion method.

\bibliography{Quasihole}

\begin{thebibliography}{52}%
\makeatletter
\providecommand \@ifxundefined [1]{%
 \@ifx{#1\undefined}
}%
\providecommand \@ifnum [1]{%
 \ifnum #1\expandafter \@firstoftwo
 \else \expandafter \@secondoftwo
 \fi
}%
\providecommand \@ifx [1]{%
 \ifx #1\expandafter \@firstoftwo
 \else \expandafter \@secondoftwo
 \fi
}%
\providecommand \natexlab [1]{#1}%
\providecommand \enquote  [1]{``#1''}%
\providecommand \bibnamefont  [1]{#1}%
\providecommand \bibfnamefont [1]{#1}%
\providecommand \citenamefont [1]{#1}%
\providecommand \href@noop [0]{\@secondoftwo}%
\providecommand \href [0]{\begingroup \@sanitize@url \@href}%
\providecommand \@href[1]{\@@startlink{#1}\@@href}%
\providecommand \@@href[1]{\endgroup#1\@@endlink}%
\providecommand \@sanitize@url [0]{\catcode `\\12\catcode `\$12\catcode
  `\&12\catcode `\#12\catcode `\^12\catcode `\_12\catcode `\%12\relax}%
\providecommand \@@startlink[1]{}%
\providecommand \@@endlink[0]{}%
\providecommand \url  [0]{\begingroup\@sanitize@url \@url }%
\providecommand \@url [1]{\endgroup\@href {#1}{\urlprefix }}%
\providecommand \urlprefix  [0]{URL }%
\providecommand \Eprint [0]{\href }%
\providecommand \doibase [0]{https://doi.org/}%
\providecommand \selectlanguage [0]{\@gobble}%
\providecommand \bibinfo  [0]{\@secondoftwo}%
\providecommand \bibfield  [0]{\@secondoftwo}%
\providecommand \translation [1]{[#1]}%
\providecommand \BibitemOpen [0]{}%
\providecommand \bibitemStop [0]{}%
\providecommand \bibitemNoStop [0]{.\EOS\space}%
\providecommand \EOS [0]{\spacefactor3000\relax}%
\providecommand \BibitemShut  [1]{\csname bibitem#1\endcsname}%
\let\auto@bib@innerbib\@empty
\bibitem [{\citenamefont {Giuliani}\ and\ \citenamefont
  {Vignale}(2008)}]{giuliani2008quantum}%
  \BibitemOpen
  \bibfield  {author} {\bibinfo {author} {\bibfnamefont {G.}~\bibnamefont
  {Giuliani}}\ and\ \bibinfo {author} {\bibfnamefont {G.}~\bibnamefont
  {Vignale}},\ }\href@noop {} {\emph {\bibinfo {title} {Quantum theory of the
  electron liquid}}}\ (\bibinfo  {publisher} {Cambridge university press},\
  \bibinfo {year} {2008})\BibitemShut {NoStop}%
\bibitem [{\citenamefont {Haldane}(1981)}]{haldane1981luttinger}%
  \BibitemOpen
  \bibfield  {author} {\bibinfo {author} {\bibfnamefont {F.}~\bibnamefont
  {Haldane}},\ }\bibfield  {title} {\bibinfo {title} {'luttinger liquid
  theory'of one-dimensional quantum fluids. i. properties of the luttinger
  model and their extension to the general 1d interacting spinless fermi gas},\
  }\href {https://iopscience.iop.org/article/10.1088/0022-3719/14/19/010/meta}
  {\bibfield  {journal} {\bibinfo  {journal} {Journal of Physics C: Solid State
  Physics}\ }\textbf {\bibinfo {volume} {14}},\ \bibinfo {pages} {2585}
  (\bibinfo {year} {1981})}\BibitemShut {NoStop}%
\bibitem [{\citenamefont {Halperin}\ \emph {et~al.}(1993)\citenamefont
  {Halperin}, \citenamefont {Lee},\ and\ \citenamefont
  {Read}}]{halperin_theory_1993}%
  \BibitemOpen
  \bibfield  {author} {\bibinfo {author} {\bibfnamefont {B.~I.}\ \bibnamefont
  {Halperin}}, \bibinfo {author} {\bibfnamefont {P.~A.}\ \bibnamefont {Lee}},\
  and\ \bibinfo {author} {\bibfnamefont {N.}~\bibnamefont {Read}},\ }\bibfield
  {title} {\bibinfo {title} {Theory of the half-filled {Landau} level},\ }\href
  {https://doi.org/10.1103/PhysRevB.47.7312} {\bibfield  {journal} {\bibinfo
  {journal} {Phys. Rev. B}\ }\textbf {\bibinfo {volume} {47}},\ \bibinfo
  {pages} {7312} (\bibinfo {year} {1993})}\BibitemShut {NoStop}%
\bibitem [{\citenamefont {Klitzing}\ \emph {et~al.}(1980)\citenamefont
  {Klitzing}, \citenamefont {Dorda},\ and\ \citenamefont
  {Pepper}}]{klitzing_new_1980}%
  \BibitemOpen
  \bibfield  {author} {\bibinfo {author} {\bibfnamefont {K.~v.}\ \bibnamefont
  {Klitzing}}, \bibinfo {author} {\bibfnamefont {G.}~\bibnamefont {Dorda}},\
  and\ \bibinfo {author} {\bibfnamefont {M.}~\bibnamefont {Pepper}},\
  }\bibfield  {title} {\bibinfo {title} {New {Method} for {High}-{Accuracy}
  {Determination} of the {Fine}-{Structure} {Constant} {Based} on {Quantized}
  {Hall} {Resistance}},\ }\href {https://doi.org/10.1103/PhysRevLett.45.494}
  {\bibfield  {journal} {\bibinfo  {journal} {Phys. Rev. Lett.}\ }\textbf
  {\bibinfo {volume} {45}},\ \bibinfo {pages} {494} (\bibinfo {year}
  {1980})}\BibitemShut {NoStop}%
\bibitem [{\citenamefont {Tsui}\ \emph {et~al.}(1982)\citenamefont {Tsui},
  \citenamefont {Stormer},\ and\ \citenamefont
  {Gossard}}]{tsui_two-dimensional_1982}%
  \BibitemOpen
  \bibfield  {author} {\bibinfo {author} {\bibfnamefont {D.~C.}\ \bibnamefont
  {Tsui}}, \bibinfo {author} {\bibfnamefont {H.~L.}\ \bibnamefont {Stormer}},\
  and\ \bibinfo {author} {\bibfnamefont {A.~C.}\ \bibnamefont {Gossard}},\
  }\bibfield  {title} {\bibinfo {title} {Two-{Dimensional} {Magnetotransport}
  in the {Extreme} {Quantum} {Limit}},\ }\href
  {https://doi.org/10.1103/PhysRevLett.48.1559} {\bibfield  {journal} {\bibinfo
   {journal} {Phys. Rev. Lett.}\ }\textbf {\bibinfo {volume} {48}},\ \bibinfo
  {pages} {1559} (\bibinfo {year} {1982})}\BibitemShut {NoStop}%
\bibitem [{\citenamefont {Anderson}(2006)}]{anderson_strange_2006}%
  \BibitemOpen
  \bibfield  {author} {\bibinfo {author} {\bibfnamefont {P.~W.}\ \bibnamefont
  {Anderson}},\ }\bibfield  {title} {{\selectlanguage {en}\bibinfo {title} {The
  ‘strange metal’ is a projected {Fermi} liquid with edge singularities}},\
  }\href {https://doi.org/10.1038/nphys388} {\bibfield  {journal} {\bibinfo
  {journal} {Nature Phys}\ }\textbf {\bibinfo {volume} {2}},\ \bibinfo {pages}
  {626} (\bibinfo {year} {2006})}\BibitemShut {NoStop}%
\bibitem [{\citenamefont {Senaratne}\ \emph {et~al.}(2022)\citenamefont
  {Senaratne}, \citenamefont {Cavazos-Cavazos}, \citenamefont {Wang},
  \citenamefont {He}, \citenamefont {Chang}, \citenamefont {Kafle},
  \citenamefont {Pu}, \citenamefont {Guan},\ and\ \citenamefont
  {Hulet}}]{senaratne_spin-charge_2022}%
  \BibitemOpen
  \bibfield  {author} {\bibinfo {author} {\bibfnamefont {R.}~\bibnamefont
  {Senaratne}}, \bibinfo {author} {\bibfnamefont {D.}~\bibnamefont
  {Cavazos-Cavazos}}, \bibinfo {author} {\bibfnamefont {S.}~\bibnamefont
  {Wang}}, \bibinfo {author} {\bibfnamefont {F.}~\bibnamefont {He}}, \bibinfo
  {author} {\bibfnamefont {Y.-T.}\ \bibnamefont {Chang}}, \bibinfo {author}
  {\bibfnamefont {A.}~\bibnamefont {Kafle}}, \bibinfo {author} {\bibfnamefont
  {H.}~\bibnamefont {Pu}}, \bibinfo {author} {\bibfnamefont {X.-W.}\
  \bibnamefont {Guan}},\ and\ \bibinfo {author} {\bibfnamefont {R.~G.}\
  \bibnamefont {Hulet}},\ }\bibfield  {title} {{\selectlanguage {en}\bibinfo
  {title} {Spin-charge separation in a one-dimensional {Fermi} gas with tunable
  interactions}},\ }\href {https://doi.org/10.1126/science.abn1719} {\bibfield
  {journal} {\bibinfo  {journal} {Science}\ }\textbf {\bibinfo {volume}
  {376}},\ \bibinfo {pages} {1305} (\bibinfo {year} {2022})}\BibitemShut
  {NoStop}%
\bibitem [{\citenamefont {Halperin}(2020)}]{halperin2020half}%
  \BibitemOpen
  \bibfield  {author} {\bibinfo {author} {\bibfnamefont {B.~I.}\ \bibnamefont
  {Halperin}},\ }\bibfield  {title} {\bibinfo {title} {The half-full landau
  level},\ }in\ \href
  {https://www.worldscientific.com.remotexs.ntu.edu.sg/doi/10.1142/9789811217494_0002}
  {\emph {\bibinfo {booktitle} {Fractional quantum Hall effects: New
  developments}}}\ (\bibinfo  {publisher} {World Scientific},\ \bibinfo {year}
  {2020})\ pp.\ \bibinfo {pages} {79--132}\BibitemShut {NoStop}%
\bibitem [{\citenamefont {Kamilla}\ \emph {et~al.}(1997)\citenamefont
  {Kamilla}, \citenamefont {Jain},\ and\ \citenamefont {Girvin}}]{Kamilla97}%
  \BibitemOpen
  \bibfield  {author} {\bibinfo {author} {\bibfnamefont {R.~K.}\ \bibnamefont
  {Kamilla}}, \bibinfo {author} {\bibfnamefont {J.~K.}\ \bibnamefont {Jain}},\
  and\ \bibinfo {author} {\bibfnamefont {S.~M.}\ \bibnamefont {Girvin}},\
  }\bibfield  {title} {\bibinfo {title} {Fermi-sea-like correlations in a
  partially filled {Landau} level},\ }\href
  {https://doi.org/10.1103/PhysRevB.56.12411} {\bibfield  {journal} {\bibinfo
  {journal} {Phys. Rev. B}\ }\textbf {\bibinfo {volume} {56}},\ \bibinfo
  {pages} {12411} (\bibinfo {year} {1997})}\BibitemShut {NoStop}%
\bibitem [{\citenamefont {Chang}(2003)}]{chang_chiral_2003}%
  \BibitemOpen
  \bibfield  {author} {\bibinfo {author} {\bibfnamefont {A.~M.}\ \bibnamefont
  {Chang}},\ }\bibfield  {title} {{\selectlanguage {en}\bibinfo {title} {Chiral
  {Luttinger} liquids at the fractional quantum {Hall} edge}},\ }\href
  {https://doi.org/10.1103/RevModPhys.75.1449} {\bibfield  {journal} {\bibinfo
  {journal} {Rev. Mod. Phys.}\ }\textbf {\bibinfo {volume} {75}},\ \bibinfo
  {pages} {1449} (\bibinfo {year} {2003})}\BibitemShut {NoStop}%
\bibitem [{\citenamefont {Balram}\ \emph {et~al.}(2015)\citenamefont {Balram},
  \citenamefont {Tőke},\ and\ \citenamefont {Jain}}]{balram_luttinger_2015}%
  \BibitemOpen
  \bibfield  {author} {\bibinfo {author} {\bibfnamefont {A.~C.}\ \bibnamefont
  {Balram}}, \bibinfo {author} {\bibfnamefont {C.}~\bibnamefont {Tőke}},\ and\
  \bibinfo {author} {\bibfnamefont {J.}~\bibnamefont {Jain}},\ }\bibfield
  {title} {{\selectlanguage {en}\bibinfo {title} {Luttinger {Theorem} for the
  {Strongly} {Correlated} {Fermi} {Liquid} of {Composite} {Fermions}}},\ }\href
  {https://doi.org/10.1103/PhysRevLett.115.186805} {\bibfield  {journal}
  {\bibinfo  {journal} {Phys. Rev. Lett.}\ }\textbf {\bibinfo {volume} {115}},\
  \bibinfo {pages} {186805} (\bibinfo {year} {2015})}\BibitemShut {NoStop}%
\bibitem [{\citenamefont {Balram}\ and\ \citenamefont
  {Jain}(2017)}]{balram_luttinger_2017}%
  \BibitemOpen
  \bibfield  {author} {\bibinfo {author} {\bibfnamefont {A.~C.}\ \bibnamefont
  {Balram}}\ and\ \bibinfo {author} {\bibfnamefont {J.~K.}\ \bibnamefont
  {Jain}},\ }\bibfield  {title} {\bibinfo {title} {Fermi wave vector for the
  partially spin-polarized composite-fermion {Fermi} sea},\ }\href
  {https://doi.org/10.1103/PhysRevB.96.235102} {\bibfield  {journal} {\bibinfo
  {journal} {Phys. Rev. B}\ }\textbf {\bibinfo {volume} {96}},\ \bibinfo
  {pages} {235102} (\bibinfo {year} {2017})}\BibitemShut {NoStop}%
\bibitem [{\citenamefont {Mitrano}\ \emph {et~al.}(2018)\citenamefont
  {Mitrano}, \citenamefont {Husain}, \citenamefont {Vig}, \citenamefont
  {Kogar}, \citenamefont {Rak}, \citenamefont {Rubeck}, \citenamefont
  {Schmalian}, \citenamefont {Uchoa}, \citenamefont {Schneeloch}, \citenamefont
  {Zhong} \emph {et~al.}}]{mitrano2018anomalous}%
  \BibitemOpen
  \bibfield  {author} {\bibinfo {author} {\bibfnamefont {M.}~\bibnamefont
  {Mitrano}}, \bibinfo {author} {\bibfnamefont {A.}~\bibnamefont {Husain}},
  \bibinfo {author} {\bibfnamefont {S.}~\bibnamefont {Vig}}, \bibinfo {author}
  {\bibfnamefont {A.}~\bibnamefont {Kogar}}, \bibinfo {author} {\bibfnamefont
  {M.}~\bibnamefont {Rak}}, \bibinfo {author} {\bibfnamefont {S.}~\bibnamefont
  {Rubeck}}, \bibinfo {author} {\bibfnamefont {J.}~\bibnamefont {Schmalian}},
  \bibinfo {author} {\bibfnamefont {B.}~\bibnamefont {Uchoa}}, \bibinfo
  {author} {\bibfnamefont {J.}~\bibnamefont {Schneeloch}}, \bibinfo {author}
  {\bibfnamefont {R.}~\bibnamefont {Zhong}}, \emph {et~al.},\ }\bibfield
  {title} {\bibinfo {title} {Anomalous density fluctuations in a strange
  metal},\ }\href {https://www.pnas.org/doi/10.1073/pnas.1721495115} {\bibfield
   {journal} {\bibinfo  {journal} {Proceedings of the National Academy of
  Sciences}\ }\textbf {\bibinfo {volume} {115}},\ \bibinfo {pages} {5392}
  (\bibinfo {year} {2018})}\BibitemShut {NoStop}%
\bibitem [{\citenamefont {Laughlin}(1983)}]{laughlin_anomalous_1983}%
  \BibitemOpen
  \bibfield  {author} {\bibinfo {author} {\bibfnamefont {R.~B.}\ \bibnamefont
  {Laughlin}},\ }\bibfield  {title} {{\selectlanguage {en}\bibinfo {title}
  {Anomalous {Quantum} {Hall} {Effect}: {An} {Incompressible} {Quantum} {Fluid}
  with {Fractionally} {Charged} {Excitations}}},\ }\href
  {https://doi.org/10.1103/PhysRevLett.50.1395} {\bibfield  {journal} {\bibinfo
   {journal} {Phys. Rev. Lett.}\ }\textbf {\bibinfo {volume} {50}},\ \bibinfo
  {pages} {1395} (\bibinfo {year} {1983})}\BibitemShut {NoStop}%
\bibitem [{\citenamefont {Arovas}\ \emph {et~al.}(1984)\citenamefont {Arovas},
  \citenamefont {Schrieffer},\ and\ \citenamefont
  {Wilczek}}]{arovas_fractional_1984}%
  \BibitemOpen
  \bibfield  {author} {\bibinfo {author} {\bibfnamefont {D.}~\bibnamefont
  {Arovas}}, \bibinfo {author} {\bibfnamefont {J.~R.}\ \bibnamefont
  {Schrieffer}},\ and\ \bibinfo {author} {\bibfnamefont {F.}~\bibnamefont
  {Wilczek}},\ }\bibfield  {title} {{\selectlanguage {en}\bibinfo {title}
  {Fractional {Statistics} and the {Quantum} {Hall} {Effect}}},\ }\href
  {https://doi.org/10.1103/PhysRevLett.53.722} {\bibfield  {journal} {\bibinfo
  {journal} {Phys. Rev. Lett.}\ }\textbf {\bibinfo {volume} {53}},\ \bibinfo
  {pages} {722} (\bibinfo {year} {1984})}\BibitemShut {NoStop}%
\bibitem [{\citenamefont {Trung}\ \emph {et~al.}(2023)\citenamefont {Trung},
  \citenamefont {Wang},\ and\ \citenamefont
  {Yang}}]{trung_spin-statistics_2023}%
  \BibitemOpen
  \bibfield  {author} {\bibinfo {author} {\bibfnamefont {H.~Q.}\ \bibnamefont
  {Trung}}, \bibinfo {author} {\bibfnamefont {Y.}~\bibnamefont {Wang}},\ and\
  \bibinfo {author} {\bibfnamefont {B.}~\bibnamefont {Yang}},\ }\bibfield
  {title} {{\selectlanguage {en}\bibinfo {title} {Spin-statistics relation and
  {Abelian} braiding phase for anyons in the fractional quantum {Hall}
  effect}},\ }\href {https://doi.org/10.1103/PhysRevB.107.L201301} {\bibfield
  {journal} {\bibinfo  {journal} {Phys. Rev. B}\ }\textbf {\bibinfo {volume}
  {107}},\ \bibinfo {pages} {L201301} (\bibinfo {year} {2023})}\BibitemShut
  {NoStop}%
\bibitem [{\citenamefont {Park}\ and\ \citenamefont
  {Haldane}(2014)}]{park_guiding-center_2014}%
  \BibitemOpen
  \bibfield  {author} {\bibinfo {author} {\bibfnamefont {Y.}~\bibnamefont
  {Park}}\ and\ \bibinfo {author} {\bibfnamefont {F.~D.~M.}\ \bibnamefont
  {Haldane}},\ }\bibfield  {title} {\bibinfo {title} {Guiding-center {Hall}
  viscosity and intrinsic dipole moment along edges of incompressible
  fractional quantum {Hall} fluids},\ }\href
  {https://doi.org/10.1103/PhysRevB.90.045123} {\bibfield  {journal} {\bibinfo
  {journal} {Phys. Rev. B}\ }\textbf {\bibinfo {volume} {90}},\ \bibinfo
  {pages} {045123} (\bibinfo {year} {2014})}\BibitemShut {NoStop}%
\bibitem [{\citenamefont {Datta}\ \emph {et~al.}(1996)\citenamefont {Datta},
  \citenamefont {Morf},\ and\ \citenamefont {Ferrari}}]{datta_edge_1996}%
  \BibitemOpen
  \bibfield  {author} {\bibinfo {author} {\bibfnamefont {N.}~\bibnamefont
  {Datta}}, \bibinfo {author} {\bibfnamefont {R.}~\bibnamefont {Morf}},\ and\
  \bibinfo {author} {\bibfnamefont {R.}~\bibnamefont {Ferrari}},\ }\bibfield
  {title} {{\selectlanguage {en}\bibinfo {title} {Edge of the {Laughlin}
  droplet}},\ }\href {https://doi.org/10.1103/PhysRevB.53.10906} {\bibfield
  {journal} {\bibinfo  {journal} {Phys. Rev. B}\ }\textbf {\bibinfo {volume}
  {53}},\ \bibinfo {pages} {10906} (\bibinfo {year} {1996})}\BibitemShut
  {NoStop}%
\bibitem [{\citenamefont {Wen}(1990)}]{wen_chiral_1990}%
  \BibitemOpen
  \bibfield  {author} {\bibinfo {author} {\bibfnamefont {X.~G.}\ \bibnamefont
  {Wen}},\ }\bibfield  {title} {{\selectlanguage {en}\bibinfo {title} {Chiral
  {Luttinger} liquid and the edge excitations in the fractional quantum {Hall}
  states}},\ }\href {https://doi.org/10.1103/PhysRevB.41.12838} {\bibfield
  {journal} {\bibinfo  {journal} {Phys. Rev. B}\ }\textbf {\bibinfo {volume}
  {41}},\ \bibinfo {pages} {12838} (\bibinfo {year} {1990})}\BibitemShut
  {NoStop}%
\bibitem [{\citenamefont {Levesque}\ \emph {et~al.}(2000)\citenamefont
  {Levesque}, \citenamefont {Weis},\ and\ \citenamefont
  {Lebowitz}}]{levesque2000charge}%
  \BibitemOpen
  \bibfield  {author} {\bibinfo {author} {\bibfnamefont {D.}~\bibnamefont
  {Levesque}}, \bibinfo {author} {\bibfnamefont {J.-J.}\ \bibnamefont {Weis}},\
  and\ \bibinfo {author} {\bibfnamefont {J.}~\bibnamefont {Lebowitz}},\
  }\bibfield  {title} {\bibinfo {title} {Charge fluctuations in the
  two-dimensional one-component plasma},\ }\href
  {https://link.springer.com/article/10.1023/A:1018643829340} {\bibfield
  {journal} {\bibinfo  {journal} {Journal of Statistical Physics}\ }\textbf
  {\bibinfo {volume} {100}},\ \bibinfo {pages} {209} (\bibinfo {year}
  {2000})}\BibitemShut {NoStop}%
\bibitem [{\citenamefont {Wiegmann}(2012)}]{wiegmann_nonlinear_2012}%
  \BibitemOpen
  \bibfield  {author} {\bibinfo {author} {\bibfnamefont {P.}~\bibnamefont
  {Wiegmann}},\ }\bibfield  {title} {{\selectlanguage {en}\bibinfo {title}
  {Nonlinear {Hydrodynamics} and {Fractionally} {Quantized} {Solitons} at the
  {Fractional} {Quantum} {Hall} {Edge}}},\ }\href
  {https://doi.org/10.1103/PhysRevLett.108.206810} {\bibfield  {journal}
  {\bibinfo  {journal} {Phys. Rev. Lett.}\ }\textbf {\bibinfo {volume} {108}},\
  \bibinfo {pages} {206810} (\bibinfo {year} {2012})}\BibitemShut {NoStop}%
\bibitem [{\citenamefont {Can}\ \emph {et~al.}(2014)\citenamefont {Can},
  \citenamefont {Forrester}, \citenamefont {Téllez},\ and\ \citenamefont
  {Wiegmann}}]{can_singular_2014}%
  \BibitemOpen
  \bibfield  {author} {\bibinfo {author} {\bibfnamefont {T.}~\bibnamefont
  {Can}}, \bibinfo {author} {\bibfnamefont {P.~J.}\ \bibnamefont {Forrester}},
  \bibinfo {author} {\bibfnamefont {G.}~\bibnamefont {Téllez}},\ and\ \bibinfo
  {author} {\bibfnamefont {P.}~\bibnamefont {Wiegmann}},\ }\bibfield  {title}
  {{\selectlanguage {en}\bibinfo {title} {Singular behavior at the edge of
  {Laughlin} states}},\ }\href {https://doi.org/10.1103/PhysRevB.89.235137}
  {\bibfield  {journal} {\bibinfo  {journal} {Phys. Rev. B}\ }\textbf {\bibinfo
  {volume} {89}},\ \bibinfo {pages} {235137} (\bibinfo {year}
  {2014})}\BibitemShut {NoStop}%
\bibitem [{\citenamefont {Fern}\ and\ \citenamefont
  {Simon}(2017)}]{fern_quantum_2017}%
  \BibitemOpen
  \bibfield  {author} {\bibinfo {author} {\bibfnamefont {R.}~\bibnamefont
  {Fern}}\ and\ \bibinfo {author} {\bibfnamefont {S.~H.}\ \bibnamefont
  {Simon}},\ }\bibfield  {title} {{\selectlanguage {en}\bibinfo {title}
  {Quantum {Hall} edges with hard confinement: {Exact} solution beyond
  {Luttinger} liquid}},\ }\href {https://doi.org/10.1103/PhysRevB.95.201108}
  {\bibfield  {journal} {\bibinfo  {journal} {Phys. Rev. B}\ }\textbf {\bibinfo
  {volume} {95}},\ \bibinfo {pages} {201108} (\bibinfo {year}
  {2017})}\BibitemShut {NoStop}%
\bibitem [{\citenamefont {Ito}\ and\ \citenamefont
  {Shibata}(2021)}]{ito_density_2021}%
  \BibitemOpen
  \bibfield  {author} {\bibinfo {author} {\bibfnamefont {T.}~\bibnamefont
  {Ito}}\ and\ \bibinfo {author} {\bibfnamefont {N.}~\bibnamefont {Shibata}},\
  }\bibfield  {title} {\bibinfo {title} {Density matrix renormalization group
  study of the $\ensuremath{\nu}=1/3$ edge states in fractional quantum {Hall}
  systems},\ }\href {https://doi.org/10.1103/PhysRevB.103.115107} {\bibfield
  {journal} {\bibinfo  {journal} {Phys. Rev. B}\ }\textbf {\bibinfo {volume}
  {103}},\ \bibinfo {pages} {115107} (\bibinfo {year} {2021})}\BibitemShut
  {NoStop}%
\bibitem [{\citenamefont {Cardoso}\ \emph {et~al.}(2021)\citenamefont
  {Cardoso}, \citenamefont {Stéphan},\ and\ \citenamefont
  {Abanov}}]{cardoso_boundary_2021}%
  \BibitemOpen
  \bibfield  {author} {\bibinfo {author} {\bibfnamefont {G.}~\bibnamefont
  {Cardoso}}, \bibinfo {author} {\bibfnamefont {J.-M.}\ \bibnamefont
  {Stéphan}},\ and\ \bibinfo {author} {\bibfnamefont {A.~G.}\ \bibnamefont
  {Abanov}},\ }\bibfield  {title} {{\selectlanguage {en}\bibinfo {title} {The
  boundary density profile of a {Coulomb} droplet. {Freezing} at the edge}},\
  }\href {https://doi.org/10.1088/1751-8121/abcab9} {\bibfield  {journal}
  {\bibinfo  {journal} {J. Phys. A: Math. Theor.}\ }\textbf {\bibinfo {volume}
  {54}},\ \bibinfo {pages} {015002} (\bibinfo {year} {2021})}\BibitemShut
  {NoStop}%
\bibitem [{\citenamefont {Yang}\ and\ \citenamefont
  {Hu}(2023)}]{yang_monte_2023}%
  \BibitemOpen
  \bibfield  {author} {\bibinfo {author} {\bibfnamefont {Y.}~\bibnamefont
  {Yang}}\ and\ \bibinfo {author} {\bibfnamefont {Z.-X.}\ \bibnamefont {Hu}},\
  }\bibfield  {title} {{\selectlanguage {en}\bibinfo {title} {Monte {Carlo}
  simulation of the topological quantities in fractional quantum {Hall}
  systems}},\ }\href {https://doi.org/10.1103/PhysRevB.107.115162} {\bibfield
  {journal} {\bibinfo  {journal} {Phys. Rev. B}\ }\textbf {\bibinfo {volume}
  {107}},\ \bibinfo {pages} {115162} (\bibinfo {year} {2023})}\BibitemShut
  {NoStop}%
\bibitem [{\citenamefont {Girvin}\ \emph {et~al.}(1986)\citenamefont {Girvin},
  \citenamefont {MacDonald},\ and\ \citenamefont
  {Platzman}}]{girvin_magneto-roton_1986}%
  \BibitemOpen
  \bibfield  {author} {\bibinfo {author} {\bibfnamefont {S.~M.}\ \bibnamefont
  {Girvin}}, \bibinfo {author} {\bibfnamefont {A.~H.}\ \bibnamefont
  {MacDonald}},\ and\ \bibinfo {author} {\bibfnamefont {P.~M.}\ \bibnamefont
  {Platzman}},\ }\bibfield  {title} {{\selectlanguage {en}\bibinfo {title}
  {Magneto-roton theory of collective excitations in the fractional quantum
  {Hall} effect}},\ }\href {https://doi.org/10.1103/PhysRevB.33.2481}
  {\bibfield  {journal} {\bibinfo  {journal} {Phys. Rev. B}\ }\textbf {\bibinfo
  {volume} {33}},\ \bibinfo {pages} {2481} (\bibinfo {year}
  {1986})}\BibitemShut {NoStop}%
\bibitem [{\citenamefont {Johri}\ \emph {et~al.}(2014)\citenamefont {Johri},
  \citenamefont {Papić}, \citenamefont {Bhatt},\ and\ \citenamefont
  {Schmitteckert}}]{johri_quasiholes_2014}%
  \BibitemOpen
  \bibfield  {author} {\bibinfo {author} {\bibfnamefont {S.}~\bibnamefont
  {Johri}}, \bibinfo {author} {\bibfnamefont {Z.}~\bibnamefont {Papić}},
  \bibinfo {author} {\bibfnamefont {R.~N.}\ \bibnamefont {Bhatt}},\ and\
  \bibinfo {author} {\bibfnamefont {P.}~\bibnamefont {Schmitteckert}},\
  }\bibfield  {title} {{\selectlanguage {en}\bibinfo {title} {Quasiholes of 1 3
  and 7 3 quantum {Hall} states: {Size} estimates via exact diagonalization and
  density-matrix renormalization group}},\ }\href
  {https://doi.org/10.1103/PhysRevB.89.115124} {\bibfield  {journal} {\bibinfo
  {journal} {Phys. Rev. B}\ }\textbf {\bibinfo {volume} {89}},\ \bibinfo
  {pages} {115124} (\bibinfo {year} {2014})}\BibitemShut {NoStop}%
\bibitem [{\citenamefont {Fulsebakke}\ \emph {et~al.}(2023)\citenamefont
  {Fulsebakke}, \citenamefont {Fremling}, \citenamefont {Moran},\ and\
  \citenamefont {Slingerland}}]{fulsebakke_parametrization_2023}%
  \BibitemOpen
  \bibfield  {author} {\bibinfo {author} {\bibfnamefont {J.}~\bibnamefont
  {Fulsebakke}}, \bibinfo {author} {\bibfnamefont {M.}~\bibnamefont
  {Fremling}}, \bibinfo {author} {\bibfnamefont {N.}~\bibnamefont {Moran}},\
  and\ \bibinfo {author} {\bibfnamefont {J.~K.}\ \bibnamefont {Slingerland}},\
  }\bibfield  {title} {{\selectlanguage {en}\bibinfo {title} {Parametrization
  and thermodynamic scaling of pair correlation functions for the fractional
  quantum {Hall} effect}},\ }\href
  {https://doi.org/10.21468/SciPostPhys.14.6.149} {\bibfield  {journal}
  {\bibinfo  {journal} {SciPost Phys.}\ }\textbf {\bibinfo {volume} {14}},\
  \bibinfo {pages} {149} (\bibinfo {year} {2023})}\BibitemShut {NoStop}%
\bibitem [{\citenamefont {Haldane}(1983)}]{haldane_fractional_1983}%
  \BibitemOpen
  \bibfield  {author} {\bibinfo {author} {\bibfnamefont {F.~D.~M.}\
  \bibnamefont {Haldane}},\ }\bibfield  {title} {{\selectlanguage {en}\bibinfo
  {title} {Fractional {Quantization} of the {Hall} {Effect}: {A} {Hierarchy} of
  {Incompressible} {Quantum} {Fluid} {States}}},\ }\href
  {https://doi.org/10.1103/PhysRevLett.51.605} {\bibfield  {journal} {\bibinfo
  {journal} {Phys. Rev. Lett.}\ }\textbf {\bibinfo {volume} {51}},\ \bibinfo
  {pages} {605} (\bibinfo {year} {1983})}\BibitemShut {NoStop}%
\bibitem [{\citenamefont {Jain}(2007)}]{jain2007composite}%
  \BibitemOpen
  \bibfield  {author} {\bibinfo {author} {\bibfnamefont {J.}~\bibnamefont
  {Jain}},\ }\href {https://books.google.com.hk/books?id=0jv9UF6UL20C} {\emph
  {\bibinfo {title} {Composite Fermions}}}\ (\bibinfo  {publisher} {Cambridge
  University Press},\ \bibinfo {year} {2007})\BibitemShut {NoStop}%
\bibitem [{\citenamefont {Ciftja}\ and\ \citenamefont
  {Wexler}(2003)}]{Ciftja03}%
  \BibitemOpen
  \bibfield  {author} {\bibinfo {author} {\bibfnamefont {O.}~\bibnamefont
  {Ciftja}}\ and\ \bibinfo {author} {\bibfnamefont {C.}~\bibnamefont
  {Wexler}},\ }\bibfield  {title} {\bibinfo {title} {{Monte} {Carlo} simulation
  method for {Laughlin}-like states in a disk geometry},\ }\href
  {https://doi.org/10.1103/PhysRevB.67.075304} {\bibfield  {journal} {\bibinfo
  {journal} {Phys. Rev. B}\ }\textbf {\bibinfo {volume} {67}},\ \bibinfo
  {pages} {075304} (\bibinfo {year} {2003})}\BibitemShut {NoStop}%
\bibitem [{\citenamefont {Balram}\ and\ \citenamefont
  {W\'ojs}(2020)}]{balram_fractional_2020}%
  \BibitemOpen
  \bibfield  {author} {\bibinfo {author} {\bibfnamefont {A.~C.}\ \bibnamefont
  {Balram}}\ and\ \bibinfo {author} {\bibfnamefont {A.}~\bibnamefont
  {W\'ojs}},\ }\bibfield  {title} {\bibinfo {title} {Fractional quantum {Hall}
  effect at $\ensuremath{\nu}=2+4/9$},\ }\href
  {https://doi.org/10.1103/PhysRevResearch.2.032035} {\bibfield  {journal}
  {\bibinfo  {journal} {Phys. Rev. Res.}\ }\textbf {\bibinfo {volume} {2}},\
  \bibinfo {pages} {032035} (\bibinfo {year} {2020})}\BibitemShut {NoStop}%
\bibitem [{\citenamefont {Dora}\ and\ \citenamefont
  {Balram}(2023)}]{dora2023competition}%
  \BibitemOpen
  \bibfield  {author} {\bibinfo {author} {\bibfnamefont {R.~K.}\ \bibnamefont
  {Dora}}\ and\ \bibinfo {author} {\bibfnamefont {A.~C.}\ \bibnamefont
  {Balram}},\ }\bibfield  {title} {\bibinfo {title} {Competition between
  fractional quantum {Hall} liquid and electron solid phases in the {Landau}
  levels of multilayer graphene},\ }\href
  {https://doi.org/10.1103/PhysRevB.108.235153} {\bibfield  {journal} {\bibinfo
   {journal} {Phys. Rev. B}\ }\textbf {\bibinfo {volume} {108}},\ \bibinfo
  {pages} {235153} (\bibinfo {year} {2023})}\BibitemShut {NoStop}%
\bibitem [{\citenamefont {Wigner}(1934)}]{Wigner34}%
  \BibitemOpen
  \bibfield  {author} {\bibinfo {author} {\bibfnamefont {E.}~\bibnamefont
  {Wigner}},\ }\bibfield  {title} {\bibinfo {title} {On the interaction of
  electrons in metals.},\ }\href@noop {} {\bibfield  {journal} {\bibinfo
  {journal} {Phys. Rev.}\ }\textbf {\bibinfo {volume} {46}},\ \bibinfo {pages}
  {1002} (\bibinfo {year} {1934})}\BibitemShut {NoStop}%
\bibitem [{\citenamefont {Thacker}(1981)}]{thacker_exact_1981}%
  \BibitemOpen
  \bibfield  {author} {\bibinfo {author} {\bibfnamefont {W.~C.}\ \bibnamefont
  {Thacker}},\ }\bibfield  {title} {{\selectlanguage {en}\bibinfo {title} {Some
  exact solutions to the nonlinear shallow-water wave equations}},\ }\href
  {https://doi.org/10.1017/S0022112081001882} {\bibfield  {journal} {\bibinfo
  {journal} {J. Fluid Mech.}\ }\textbf {\bibinfo {volume} {107}},\ \bibinfo
  {pages} {499} (\bibinfo {year} {1981})}\BibitemShut {NoStop}%
\bibitem [{\citenamefont {Sampson}(2008)}]{sampson2008some}%
  \BibitemOpen
  \bibfield  {author} {\bibinfo {author} {\bibfnamefont {J.}~\bibnamefont
  {Sampson}},\ }\href@noop {} {\emph {\bibinfo {title} {Some solutions of the
  shallow water wave equations}}}\ (\bibinfo  {publisher} {Swinburne University
  of Technology, Faculty of Engineering and Industrial Sciences},\ \bibinfo
  {year} {2008})\BibitemShut {NoStop}%
\bibitem [{\citenamefont {Balram}\ \emph
  {et~al.}(2013{\natexlab{a}})\citenamefont {Balram}, \citenamefont {Wu},
  \citenamefont {Sreejith}, \citenamefont {Wójs},\ and\ \citenamefont
  {Jain}}]{balram_role_2013}%
  \BibitemOpen
  \bibfield  {author} {\bibinfo {author} {\bibfnamefont {A.~C.}\ \bibnamefont
  {Balram}}, \bibinfo {author} {\bibfnamefont {Y.-H.}\ \bibnamefont {Wu}},
  \bibinfo {author} {\bibfnamefont {G.~J.}\ \bibnamefont {Sreejith}}, \bibinfo
  {author} {\bibfnamefont {A.}~\bibnamefont {Wójs}},\ and\ \bibinfo {author}
  {\bibfnamefont {J.~K.}\ \bibnamefont {Jain}},\ }\bibfield  {title}
  {{\selectlanguage {en}\bibinfo {title} {Role of {Exciton} {Screening} in the
  7 / 3 {Fractional} {Quantum} {Hall} {Effect}}},\ }\href
  {https://doi.org/10.1103/PhysRevLett.110.186801} {\bibfield  {journal}
  {\bibinfo  {journal} {Phys. Rev. Lett.}\ }\textbf {\bibinfo {volume} {110}},\
  \bibinfo {pages} {186801} (\bibinfo {year} {2013}{\natexlab{a}})}\BibitemShut
  {NoStop}%
\bibitem [{\citenamefont {Balram}\ \emph
  {et~al.}(2013{\natexlab{b}})\citenamefont {Balram}, \citenamefont {W\'ojs},\
  and\ \citenamefont {Jain}}]{Balram13}%
  \BibitemOpen
  \bibfield  {author} {\bibinfo {author} {\bibfnamefont {A.~C.}\ \bibnamefont
  {Balram}}, \bibinfo {author} {\bibfnamefont {A.}~\bibnamefont {W\'ojs}},\
  and\ \bibinfo {author} {\bibfnamefont {J.~K.}\ \bibnamefont {Jain}},\
  }\bibfield  {title} {\bibinfo {title} {State counting for excited bands of
  the fractional quantum {Hall} effect: Exclusion rules for bound excitons},\
  }\href {https://doi.org/10.1103/PhysRevB.88.205312} {\bibfield  {journal}
  {\bibinfo  {journal} {Phys. Rev. B}\ }\textbf {\bibinfo {volume} {88}},\
  \bibinfo {pages} {205312} (\bibinfo {year} {2013}{\natexlab{b}})}\BibitemShut
  {NoStop}%
\bibitem [{\citenamefont {Yang}\ \emph {et~al.}(2012)\citenamefont {Yang},
  \citenamefont {Hu}, \citenamefont {Papić},\ and\ \citenamefont
  {Haldane}}]{yang_model_2012}%
  \BibitemOpen
  \bibfield  {author} {\bibinfo {author} {\bibfnamefont {B.}~\bibnamefont
  {Yang}}, \bibinfo {author} {\bibfnamefont {Z.-X.}\ \bibnamefont {Hu}},
  \bibinfo {author} {\bibfnamefont {Z.}~\bibnamefont {Papić}},\ and\ \bibinfo
  {author} {\bibfnamefont {F.~D.~M.}\ \bibnamefont {Haldane}},\ }\bibfield
  {title} {{\selectlanguage {en}\bibinfo {title} {Model {Wave} {Functions} for
  the {Collective} {Modes} and the {Magnetoroton} {Theory} of the {Fractional}
  {Quantum} {Hall} {Effect}}},\ }\href
  {https://doi.org/10.1103/PhysRevLett.108.256807} {\bibfield  {journal}
  {\bibinfo  {journal} {Phys. Rev. Lett.}\ }\textbf {\bibinfo {volume} {108}},\
  \bibinfo {pages} {256807} (\bibinfo {year} {2012})}\BibitemShut {NoStop}%
\bibitem [{\citenamefont {Mandal}\ and\ \citenamefont
  {Jain}(2001)}]{mandal2001universal}%
  \BibitemOpen
  \bibfield  {author} {\bibinfo {author} {\bibfnamefont {S.~S.}\ \bibnamefont
  {Mandal}}\ and\ \bibinfo {author} {\bibfnamefont {J.}~\bibnamefont {Jain}},\
  }\bibfield  {title} {\bibinfo {title} {How universal is the
  fractional-quantum-hall edge luttinger liquid?},\ }\href
  {https://www.sciencedirect.com/science/article/abs/pii/S0038109801001569}
  {\bibfield  {journal} {\bibinfo  {journal} {Solid state communications}\
  }\textbf {\bibinfo {volume} {118}},\ \bibinfo {pages} {503} (\bibinfo {year}
  {2001})}\BibitemShut {NoStop}%
\bibitem [{\citenamefont {Wan}\ \emph {et~al.}(2005)\citenamefont {Wan},
  \citenamefont {Evers},\ and\ \citenamefont {Rezayi}}]{wan_universality_2005}%
  \BibitemOpen
  \bibfield  {author} {\bibinfo {author} {\bibfnamefont {X.}~\bibnamefont
  {Wan}}, \bibinfo {author} {\bibfnamefont {F.}~\bibnamefont {Evers}},\ and\
  \bibinfo {author} {\bibfnamefont {E.~H.}\ \bibnamefont {Rezayi}},\ }\bibfield
   {title} {{\selectlanguage {en}\bibinfo {title} {Universality of the
  {Edge}-{Tunneling} {Exponent} of {Fractional} {Quantum} {Hall} {Liquids}}},\
  }\href {https://doi.org/10.1103/PhysRevLett.94.166804} {\bibfield  {journal}
  {\bibinfo  {journal} {Phys. Rev. Lett.}\ }\textbf {\bibinfo {volume} {94}},\
  \bibinfo {pages} {166804} (\bibinfo {year} {2005})}\BibitemShut {NoStop}%
\bibitem [{\citenamefont {Yang}(2021)}]{yang_statistical_2021}%
  \BibitemOpen
  \bibfield  {author} {\bibinfo {author} {\bibfnamefont {B.}~\bibnamefont
  {Yang}},\ }\bibfield  {title} {{\selectlanguage {en}\bibinfo {title}
  {Statistical {Interactions} and {Boson}-{Anyon} {Duality} in {Fractional}
  {Quantum} {Hall} {Fluids}}},\ }\href
  {https://doi.org/10.1103/PhysRevLett.127.126406} {\bibfield  {journal}
  {\bibinfo  {journal} {Phys. Rev. Lett.}\ }\textbf {\bibinfo {volume} {127}},\
  \bibinfo {pages} {126406} (\bibinfo {year} {2021})}\BibitemShut {NoStop}%
\bibitem [{\citenamefont {Moore}\ and\ \citenamefont
  {Read}(1991)}]{moore_nonabelions_1991}%
  \BibitemOpen
  \bibfield  {author} {\bibinfo {author} {\bibfnamefont {G.}~\bibnamefont
  {Moore}}\ and\ \bibinfo {author} {\bibfnamefont {N.}~\bibnamefont {Read}},\
  }\bibfield  {title} {{\selectlanguage {en}\bibinfo {title} {Nonabelions in
  the fractional quantum hall effect}},\ }\href
  {https://doi.org/10.1016/0550-3213(91)90407-O} {\bibfield  {journal}
  {\bibinfo  {journal} {Nuclear Physics B}\ }\textbf {\bibinfo {volume}
  {360}},\ \bibinfo {pages} {362} (\bibinfo {year} {1991})}\BibitemShut
  {NoStop}%
\bibitem [{\citenamefont {Read}\ and\ \citenamefont
  {Green}(2000)}]{read_paired_2000}%
  \BibitemOpen
  \bibfield  {author} {\bibinfo {author} {\bibfnamefont {N.}~\bibnamefont
  {Read}}\ and\ \bibinfo {author} {\bibfnamefont {D.}~\bibnamefont {Green}},\
  }\bibfield  {title} {{\selectlanguage {en}\bibinfo {title} {Paired states of
  fermions in two dimensions with breaking of parity and time-reversal
  symmetries and the fractional quantum {Hall} effect}},\ }\href
  {https://doi.org/10.1103/PhysRevB.61.10267} {\bibfield  {journal} {\bibinfo
  {journal} {Phys. Rev. B}\ }\textbf {\bibinfo {volume} {61}},\ \bibinfo
  {pages} {10267} (\bibinfo {year} {2000})}\BibitemShut {NoStop}%
\bibitem [{\citenamefont {Simon}\ \emph {et~al.}(2007)\citenamefont {Simon},
  \citenamefont {Rezayi}, \citenamefont {Cooper},\ and\ \citenamefont
  {Berdnikov}}]{Simon07b}%
  \BibitemOpen
  \bibfield  {author} {\bibinfo {author} {\bibfnamefont {S.~H.}\ \bibnamefont
  {Simon}}, \bibinfo {author} {\bibfnamefont {E.~H.}\ \bibnamefont {Rezayi}},
  \bibinfo {author} {\bibfnamefont {N.~R.}\ \bibnamefont {Cooper}},\ and\
  \bibinfo {author} {\bibfnamefont {I.}~\bibnamefont {Berdnikov}},\ }\bibfield
  {title} {\bibinfo {title} {Construction of a paired wave function for
  spinless electrons at filling fraction $\nu=2/5$},\ }\href
  {https://doi.org/10.1103/PhysRevB.75.075317} {\bibfield  {journal} {\bibinfo
  {journal} {Phys. Rev. B}\ }\textbf {\bibinfo {volume} {75}},\ \bibinfo
  {pages} {075317} (\bibinfo {year} {2007})}\BibitemShut {NoStop}%
\bibitem [{\citenamefont {Read}\ and\ \citenamefont
  {Rezayi}(1999)}]{read_beyond_1999}%
  \BibitemOpen
  \bibfield  {author} {\bibinfo {author} {\bibfnamefont {N.}~\bibnamefont
  {Read}}\ and\ \bibinfo {author} {\bibfnamefont {E.}~\bibnamefont {Rezayi}},\
  }\bibfield  {title} {{\selectlanguage {en}\bibinfo {title} {Beyond paired
  quantum {Hall} states: {Parafermions} and incompressible states in the first
  excited {Landau} level}},\ }\href {https://doi.org/10.1103/PhysRevB.59.8084}
  {\bibfield  {journal} {\bibinfo  {journal} {Phys. Rev. B}\ }\textbf {\bibinfo
  {volume} {59}},\ \bibinfo {pages} {8084} (\bibinfo {year}
  {1999})}\BibitemShut {NoStop}%
\bibitem [{\citenamefont {Jain}(1989)}]{Jain89}%
  \BibitemOpen
  \bibfield  {author} {\bibinfo {author} {\bibfnamefont {J.~K.}\ \bibnamefont
  {Jain}},\ }\bibfield  {title} {\bibinfo {title} {Composite-fermion approach
  for the fractional quantum {Hall} effect},\ }\href
  {https://doi.org/10.1103/PhysRevLett.63.199} {\bibfield  {journal} {\bibinfo
  {journal} {Phys. Rev. Lett.}\ }\textbf {\bibinfo {volume} {63}},\ \bibinfo
  {pages} {199} (\bibinfo {year} {1989})}\BibitemShut {NoStop}%
\bibitem [{\citenamefont {Zhao}\ \emph {et~al.}(2011)\citenamefont {Zhao},
  \citenamefont {Sheng},\ and\ \citenamefont {Haldane}}]{zhao_fractional_2011}%
  \BibitemOpen
  \bibfield  {author} {\bibinfo {author} {\bibfnamefont {J.}~\bibnamefont
  {Zhao}}, \bibinfo {author} {\bibfnamefont {D.~N.}\ \bibnamefont {Sheng}},\
  and\ \bibinfo {author} {\bibfnamefont {F.~D.~M.}\ \bibnamefont {Haldane}},\
  }\bibfield  {title} {{\selectlanguage {en}\bibinfo {title} {Fractional
  quantum {Hall} states at 1 3 and 5 2 filling: {Density}-matrix
  renormalization group calculations}},\ }\href
  {https://doi.org/10.1103/PhysRevB.83.195135} {\bibfield  {journal} {\bibinfo
  {journal} {Phys. Rev. B}\ }\textbf {\bibinfo {volume} {83}},\ \bibinfo
  {pages} {195135} (\bibinfo {year} {2011})}\BibitemShut {NoStop}%
\bibitem [{\citenamefont {Mitra}\ and\ \citenamefont
  {MacDonald}(1993)}]{mitra_angular-momentum-state_1993}%
  \BibitemOpen
  \bibfield  {author} {\bibinfo {author} {\bibfnamefont {S.}~\bibnamefont
  {Mitra}}\ and\ \bibinfo {author} {\bibfnamefont {A.~H.}\ \bibnamefont
  {MacDonald}},\ }\bibfield  {title} {{\selectlanguage {en}\bibinfo {title}
  {Angular-momentum-state occupation-number distribution function of the
  {Laughlin} droplet}},\ }\href {https://doi.org/10.1103/PhysRevB.48.2005}
  {\bibfield  {journal} {\bibinfo  {journal} {Phys. Rev. B}\ }\textbf {\bibinfo
  {volume} {48}},\ \bibinfo {pages} {2005} (\bibinfo {year}
  {1993})}\BibitemShut {NoStop}%
\bibitem [{dia()}]{diagham}%
  \BibitemOpen
  \href@noop {} {}\bibinfo {note} {Diag{H}am,
  \url{https://www.nick-ux.org/diagham}}\BibitemShut {NoStop}%
\bibitem [{\citenamefont {Bernevig}\ and\ \citenamefont
  {Haldane}(2008)}]{bernevig_model_2008}%
  \BibitemOpen
  \bibfield  {author} {\bibinfo {author} {\bibfnamefont {B.~A.}\ \bibnamefont
  {Bernevig}}\ and\ \bibinfo {author} {\bibfnamefont {F.~D.~M.}\ \bibnamefont
  {Haldane}},\ }\bibfield  {title} {\bibinfo {title} {Model {Fractional}
  {Quantum} {Hall} {States} and {Jack} {Polynomials}},\ }\href
  {https://doi.org/10.1103/PhysRevLett.100.246802} {\bibfield  {journal}
  {\bibinfo  {journal} {Phys. Rev. Lett.}\ }\textbf {\bibinfo {volume} {100}},\
  \bibinfo {pages} {246802} (\bibinfo {year} {2008})}\BibitemShut {NoStop}%
\end{thebibliography}%

\end{document}